\documentclass[preprint2]{aastex}

\shorttitle{Orbits of ``fast'' visual binaries}
\shortauthors{Tokovinin}

\shorttitle{Orbits of ``fast'' visual binaries}
\shortauthors{Tokovinin}

\begin{document}

\renewcommand{\topfraction}{1.0}
\renewcommand{\bottomfraction}{1.0}
\renewcommand{\textfraction}{0.0}

\title{Speckle interferometry and orbits of ``fast'' visual binaries}

\author{Andrei Tokovinin}
\affil{Cerro Tololo Inter-American Observatory, Casilla 603, La Serena, Chile}
\email{atokovinin@ctio.noao.edu}

\begin{abstract}
Results of  speckle observations at  the 4.1-m SOAR telescope  in 2012
(158 measures of 121 systems, 27 non-resolutions) are reported. The aim
is to follow  fast orbital motion of recently  discovered or neglected
close binaries  and sub-systems.  Here  8 previously known  orbits are
defined better, two  more are completely revised, and  five orbits are
computed  for  the first  time.   Using  differential photometry  from
{\it Hipparcos}  or  speckle and  the  standard  relation  between mass  and
absolute  magnitude, the component's  masses and  dynamical parallaxes
are  estimated for all  15 systems  with new  or updated  orbits.  Two
astrometric binaries  HIP 54214  and 56245 are  resolved here  for the
first time,  another 8 are measured.  We  highlight several unresolved
pairs  that may  actually be  single despite    multiple historic
measures,  such as  104~Tau  and f~Pup  AB.   Continued monitoring  is
needed to understand those enigmatic cases.
\end{abstract}


\section{Introduction}

Close  binaries   with  fast   orbital  motion  resolved   by  speckle
interferometry  and  adaptive  optics  require  frequent  measures  to
compute their  orbits. This shifts  the main problem of  visual orbits
from  insufficiently  long  time  coverage, as  typical  for  binaries
studied  in the  past two  centuries, to  sparse time  sampling.  Many
close  pairs discovered by  W.~S.~Finsen in  the 1960s  have completed
several revolutions,  but their  orbits still remain  undetermined for
the  lack  of data.   This  is also  true  for  speckle and  {\it Hipparcos}
binaries discovered in  the 1980s and 1990s but  not followed further.
We try to address this  issue here by re-visiting close pairs resolved
recently and other binaries in need of follow-up.

Knowledge of binary-star orbits is  of fundamental value to many areas
of astronomy.  They provide  direct measurements of stellar masses and
distances,  inform  us on  the  processes  of  star formation  through
statistics  of  orbital  elements,  and  allow  dynamical  studies  of
multiple  stellar systems,  circumstellar
matter  \citep{Kennedy2012}, and  planets \citep{Roberts2011}.  A large
fraction  of  visual  binaries  are  late-type  stars  within  100\,pc
amenable to search of exo-planets.  The current orbit catalog contains
some poor or  wrong orbital solutions based on  insufficient data. The
only  way to  improve the  situation is  by getting  new  measures and
revising those orbits.

Data  on  binary-star  measures   and  orbits  are  collected  by  the
Washington   Double   Star   Catalog,   WDS   \citep{WDS}\footnote{See
  current version at 
  \url{http://ad.usno.navy.mil/wds/}} and associated archives such as
the  {\em  4-th  Catalog  of Interferometric  Measurements  of  Binary
  Stars},   INT4\footnote{\url{http://ad.usno.navy.mil/wds/int4.html}},
and the {\it 6$^{th}$ Orbit Catalog of Orbits of Visual Binary Stars},
VB6 \citep{VB6}.\footnote{\url{http://ad.usno.navy.mil/wds/orb6.html}}
These resources are extensively used here.

This paper  continues series of  speckle interfero\-metry observations
published  by   \citet[][hereafter  TMH10]{TMH10},  \citet[][hereafter
  SAM09]{SAM09}, and  \citet[][HTM12]{HTM12}.  We used  same equipment
and  data  reduction methods.   

Section~\ref{sec:data}  recalls the  observing technique  and presents
new  measures and  non-resolutions.   Updated and  new  orbits for  15
systems are  given in Section~\ref{sec:orb}, with  estimates of masses
and  dynamical  parallaxes and  brief  comments  on  each system.   In
Section~\ref{sec:other}  we draw attention  to two  particular groups:
resolved pairs with  astrometric accelerations and unresolved binaries
which     are    either     false     discoveries    or     enigmatic.
Section~\ref{sec:concl} summarizes the results.


\section{New speckle measures}
\label{sec:data}

\subsection{Instrument and observing method}

The   observations  reported   here  were   obtained  with   the  {\it
  high-resolution camera} (HRCam) -- a fast imager designed to work at
the  4.1-m  SOAR  telescope,  either  with the  SOAR  Adaptive  Module
\citep[SAM,][]{SAM}  or as  a  stand-alone instrument.   The HRCam  is
described by \citet{TC08}.  The  first series of measurements in THM10
used HRCam installed at the infrared  port of SOAR.  In 2009 the HRCam
worked  at  SAM  during   its  first-light  tests  and  produced  some
binary-star measures published in SAM09.   It was further used in this
mode in HTM12 and here.

The  HRCam receives light  through the  SAM instrument,  including its
deformable  mirror.   However,  the  AO system  was  not  compensating
turbulence during these observations. It was tuned for the ultraviolet
laser guide star, while all visible light was sent to HRCam.
The  atmospheric dispersion  was  compensated by  two rotating  prisms
inside SAM.  The filter transmission curves are given in the instrument
manual\footnote{
  http://www.ctio.noao.edu/new/Telescopes/SOAR/Instru\-ments/SAM/archive/hrcaminst.pdf}.
We  used  mostly  the  Str\"omgren  $y$ filter  (543/22\,nm)  and  the
near-infrared $I$ filter.

The  pixel scale  of HRCam  is 15.23\,mas.   Observation of  an object
consists in  accumulation of  400 frames of  200$\times$200 pixels  each with
exposure time  of 20\,ms  or shorter.  Frames  of 400$\times$400  pixels were
recorded for pairs with  separation larger than 1\farcs5.  Each object
was  recorded   twice  and  these  two  image   cubes  were  processed
independently.  Parameters  of resolved binary and  triple systems are
determined by fitting  a model to the power  spectrum, as explained in
TMH10.

Speckle interferometry of binary stars was carried out serendipitously
during five engineering runs of the SAM instrument, from December 2011
to May  2012. These observations  used short time periods  where the
SAM  engineering could  not be  pursued either  for  technical reasons
(hardware failures) or  for poor observing conditions (transparent clouds,
bad seeing). Cumulative time used  by these observations is about
one night.

We calibrated the  transfer optics of the SAM  instrument by imaging a
single-mode fiber at the telescope  focus. The fiber was translated by
a  micrometer   stage,  allowing  to   accurately  determine  detector
orientation relative  to the instrument  frame and the  physical pixel
scale in  microns.  We rely on  the SOAR telescope  pointing model and
its mechanics to ensure correct  orientation of the SAM focal plane on
the  sky  (which  however  was  not checked  independently  here)  and
constant  scale.   Therefore,  we  refer  the position  angle  to  the
telescope  and  use  the same  pixel  scale  as  in TMH10  and  HTM12.
Observations  of control  wide  binaries indicate  that  there are  no
detectable calibration errors at the level of $<$1\% in separation and
$1^\circ$ in  angle. If,  in the future,  the orbital motion  of those
wide binaries becomes known with a high accuracy, the present data can
be re-calibrated {\em post factum}.

\subsection{Data tables}

Table~1  lists   158  measures  of  121  resolved   binary  stars  and
sub-systems, including two new pairs.  Its columns contain (1) the WDS
designation, (2) the ``discoverer designation'' as adopted in the WDS,
(3) an alternative name, mostly  from the {\it Hipparcos} catalog, (4)
Besselian epoch  of observation, (5) filter, (6)  number of individual
data cubes,  (7,8) position angle $\theta$ in  degrees (not precessed)
and   internal  measurement  error   in  tangential   direction  $\rho
\sigma_{\theta}$ in  mas, (9,10)  separation $\rho$ in  arcseconds and
its  internal  error  $\sigma_{\rho}$   in  mas,  and  (11)  magnitude
difference $\Delta m$. An asterisk follows the value if $\Delta m$ and
the  true  quadrant are  determined  from  the resolved  long-exposure
image; a  colon indicates that  the data are  noisy and $\Delta  m$ is
likely over-estimated (see TMH10 for details).  Note that in the cases
of multiple stars, the positions  and photometry refer to the pairings
between individual stars, not with photo-centers of sub-systems.

For  stars with known  orbital elements,  columns (12--14)  of Table~1
list the residuals to the ephemeris position and the code of reference
to  the  orbit  adopted  in   VB6.\footnote{See  
\url{http://ad.usno.navy.mil/wds/orb6/wdsref.html}}    References  to
the orbits revised  here are preceded by asterisk;  large residuals to
those orbits show why the revisions were needed.

Table~2 contains  the data on 27  unresolved stars, some  of which are
listed   as  binaries   in  the   WDS  or   resolved  here   in  other
filters. Columns (1) through (6)  are the same as in Table~1, although
Column (2)  also includes other  names for objects  without discoverer
designations.  Columns  (7,8) give  the  $5  \sigma$ detection  limits
$\Delta m_5$  at $0\farcs15$ and  $1''$ separations determined  by the
procedure  described  in TMH10.   When  two  or  more data  cubes  are
processed, the best detection limits are listed.

\subsection{New pairs}
\label{sec:new}

{\bf   11056$-$1105   =   HIP  54214.}    \citet{Gon2000}   discovered
photo-center  motion  with  a  30\,yr  period  and a large  amplitude  of
0\farcs2, but did not not derive  the full set of orbital elements.  The
faint companion at 0\farcs6 and $60^\circ$ is resolved here in the $I$
band, but  not in $y$ ($\Delta  y > 6^m$). Its  position angle roughly
matches the plots of that paper.   This is an F0V star with fast axial
rotation and  the {\it Hipparcos}-2  \citep{HIP2} parallax $\pi_{\rm  HIP} =
  16.75\pm  0.34$\,mas. The  projected separation  of 35\,AU  and 30\,yr
  period hint  at a large mass sum.  \citet{Gon2000}  suggested that the
  actual parallax is about  18\,mas and that the astrometric companion
  is  massive.  Yearly  observations  will be  ideal  to follow  this
  interesting system.

{\bf 11318$-$2047  = HIP  56245 =  HR 440.} A  new faint  companion at
1\farcs05 is  found here.  This  is a $\Delta \mu$  astrometric binary
according  to  \citet{MK05},  spectral  type  F8V,  $\pi_{\rm  HIP}  =
25.98\pm  0.34$\,mas.   The  projected  separation of  40\,AU  implies
orbital  period on  the order  of $\sim  200$\,yr.  The  companion with
$\Delta  I =4.9$  must be  brighter in  the $K$ band,  but it  was not
detected by \citet{Boden05} when this  star served as a calibrator for
interferometry.

\section{Updated and new orbits}
\label{sec:orb}

In this  Section, we derive corrected  or first orbits  for some pairs
observed here.  Although calculation of orbital elements is accessible
to anyone  with a computer,  it is still  a challenging task  when the
measures are  scarce and their interpretation  is ambiguous (erroneous
measures  or quadrant  flips).   Additional help  is  provided by  the
availability  of {\it Hipparcos}  parallaxes, allowing  to  reject tentative
orbits with  improbably large or small  mass sum.  On  the other hand,
motion in  a visual orbit  affects {\it Hipparcos} reductions and  should be
included in the astrometric  solution whenever possible; otherwise the
parallax and proper motion can be biased \citep{Shat98, Sod1999}.

The focus here is on  fast-moving pairs where new observations allow a
substantial progress,  as in HTM12  where 42 orbits were  computed. We
refrain from correcting orbits  with large current separations.  Of 15
orbits presented  in Table~3, eight  are corrections of  prior orbits,
two are drastic  revisions, five are new.  Final  orbital elements are
obtained by least-squares  fitting with weights inversely proportional
to the squares of observational  errors. The errors of visual measures
are assumed to be  0\farcs05, speckle interferometry at 4-m telescopes
is assigned  errors of  2\,mas, with few  exceptions such as uncertain
measures marked by colons in INT4 and obvious outliers. 
The much  larger weight of speckle measures  enforces their good
fit to the orbit. For  some preliminary orbits where the least-squares
fits  did  not converge,  we  fixed one  or  more  elements (marked  by
asterisk  instead of formal  error).  Considering  that errors  of the
input  data do  not  obey the  Gaussian  statistics, formally  derived
errors  of  orbital  elements  and goodness-of-fit  criteria  such  as
$\chi^2$  should  be  taken  with reservation,  as  order-of-magnitude
estimates  at best.   Table~3 also  gives orbital  grades in  the system
adopted by VB6 (1 -- definitive, 4 -- preliminary).

Figures 1  to 4 present orbits in  the plane of the  sky, in standard
orientation (North  up, East left)  with scale in  arcseconds. The
primary components at coordinate center  are marked by large asterisks.
The orbits  are plotted  in full  line, the prior  orbits in  dashed line
where  appropriate. The  measures  (empty squares  for visual,  filled
squares  for  speckle)  are   connected  to  their  positions  on  the
orbit. Non-resolutions are shown by connecting  predicted positions
of the secondary to the coordinate origin. Some dates of speckle
measures are shown. 

Table~4   lists   astrophysical   parameters   of   pairs   with   new
orbits. Columns (1) and (2)  repeat the WDS and HIP identifies, column
(3)  lists  the  trigonometric   parallax  and  its  error  from  
\citet{HIP2}.   The spectral type  in column  (4) and  combined visual
magnitude  $V$ in  column (5)  are  taken from  SIMBAD, the  magnitude
difference in the {\it Hipparcos} band  $\Delta$Hp in column (6) is compared
to    $\Delta  y$    from  speckle photometry  in  column
(7). When  several   measures of $\Delta  y$ are  available, we
select the best one (widest separation) and round it  to the nearest $0\fm1$.

The last  four columns  of Table~4 recall  the orbital period  $P$ and
give estimates of  component masses $M_1$ and $M_2$  and the dynamical
parallax $\pi_{\rm  dyn}$. Individual magnitude of  the components are
computed  from   $V$  and   $\Delta$Hp.   When  the   {\it  Hipparcos}
differential photometry is missing or considered unreliable (marked by
colons), $\Delta  y$ is  used instead.  Masses  of the  components are
found   from  the  standard   relation  with   absolute  $V$-magnitude
\citep{Lang} using  those magnitudes and the  {\it Hipparcos} distance
modulus.  Then  the dynamical parallax  is computed from the  mass sum
and orbital  elements $P$,  $a$.  With this  parallax, the  masses are
estimated  again and  the process  is iterated  to  convergence. These
estimates of  mass and parallax  based on standard relations  for main
sequence stars should not be mistaken for direct measurements, but can
be useful for statistics; no  meaningful errors can be assigned.  When
the  dynamical  and {\it  Hipparcos}  parallaxes  match  and the  mass
estimates correspond  to the  spectral type, it  is a  good indication
that the data are mutually consistent.

\begin{figure*}
\epsscale{1.6}
\plotone{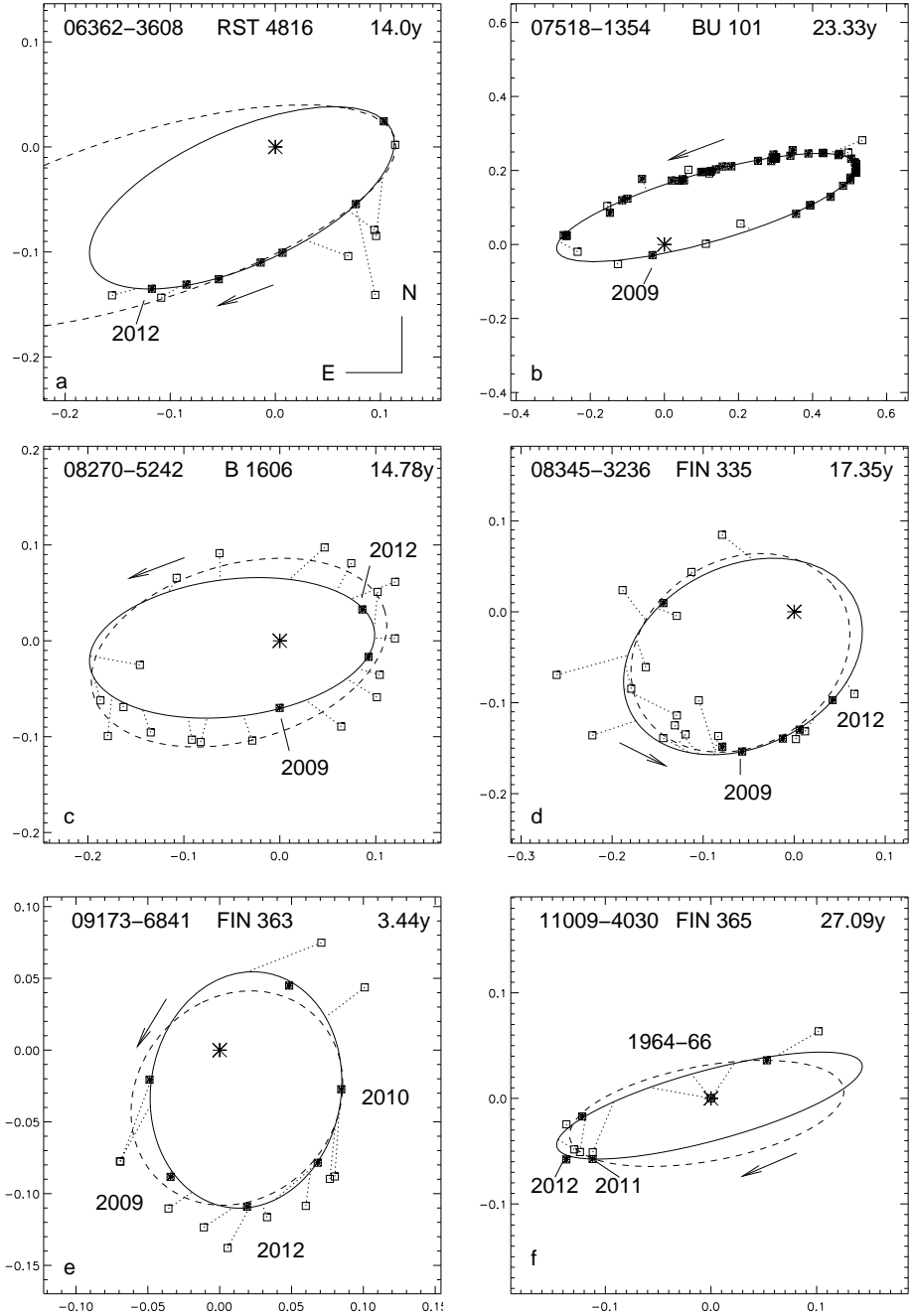}
\caption{New orbits (I). See text.
\label{fig:1} }
\end{figure*}

\begin{figure*}
\epsscale{1.6}
\plotone{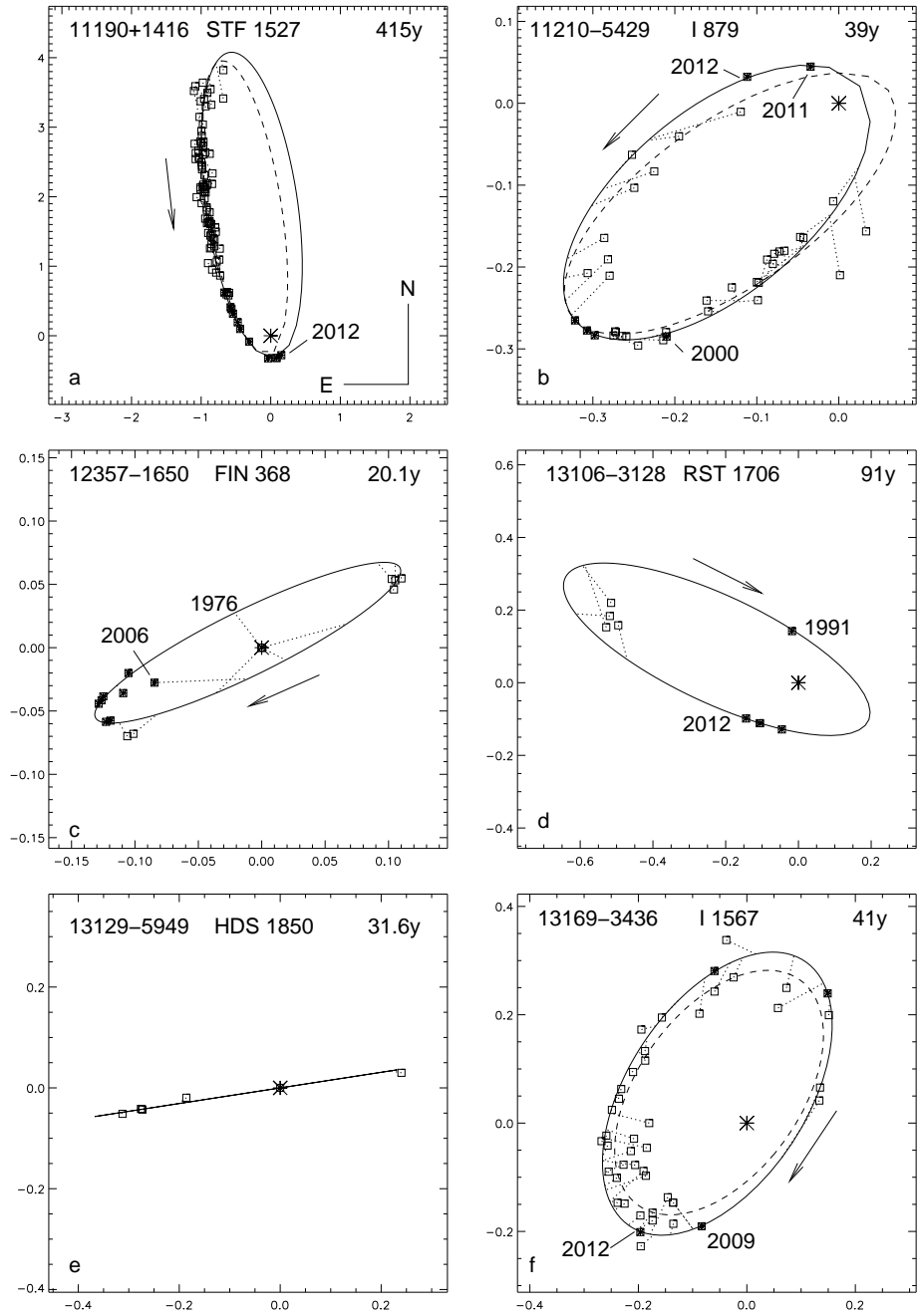}
\caption{New orbits (II).
\label{fig:2} }
\end{figure*}

\begin{figure*}
\epsscale{1.8}
\plotone{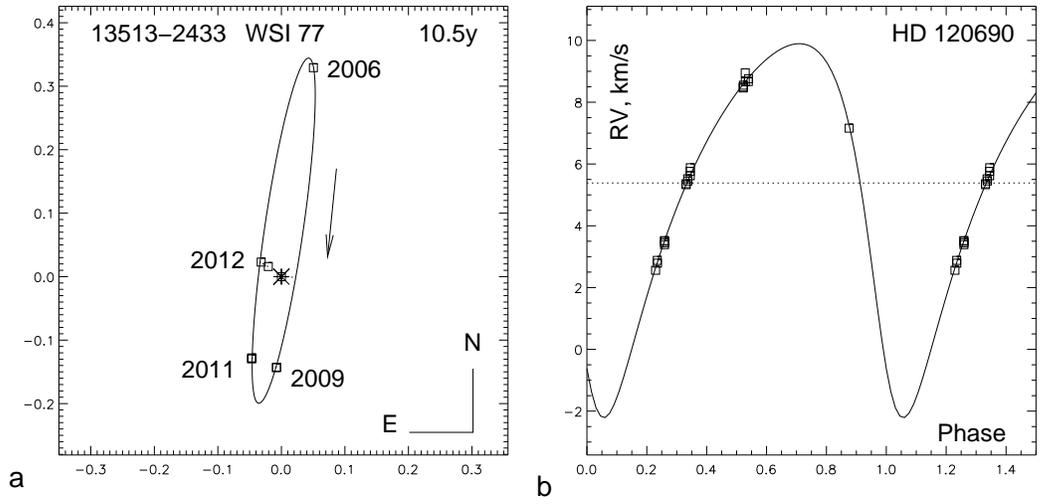}
\caption{New orbits (III): the combined orbit of WSI~77 = HD~120690. Speckle measures are plotted 
on the left (a), radial velocities on the right (b). 
\label{fig:WSI77} }
\end{figure*}

\begin{figure*}
\epsscale{1.8}
\plotone{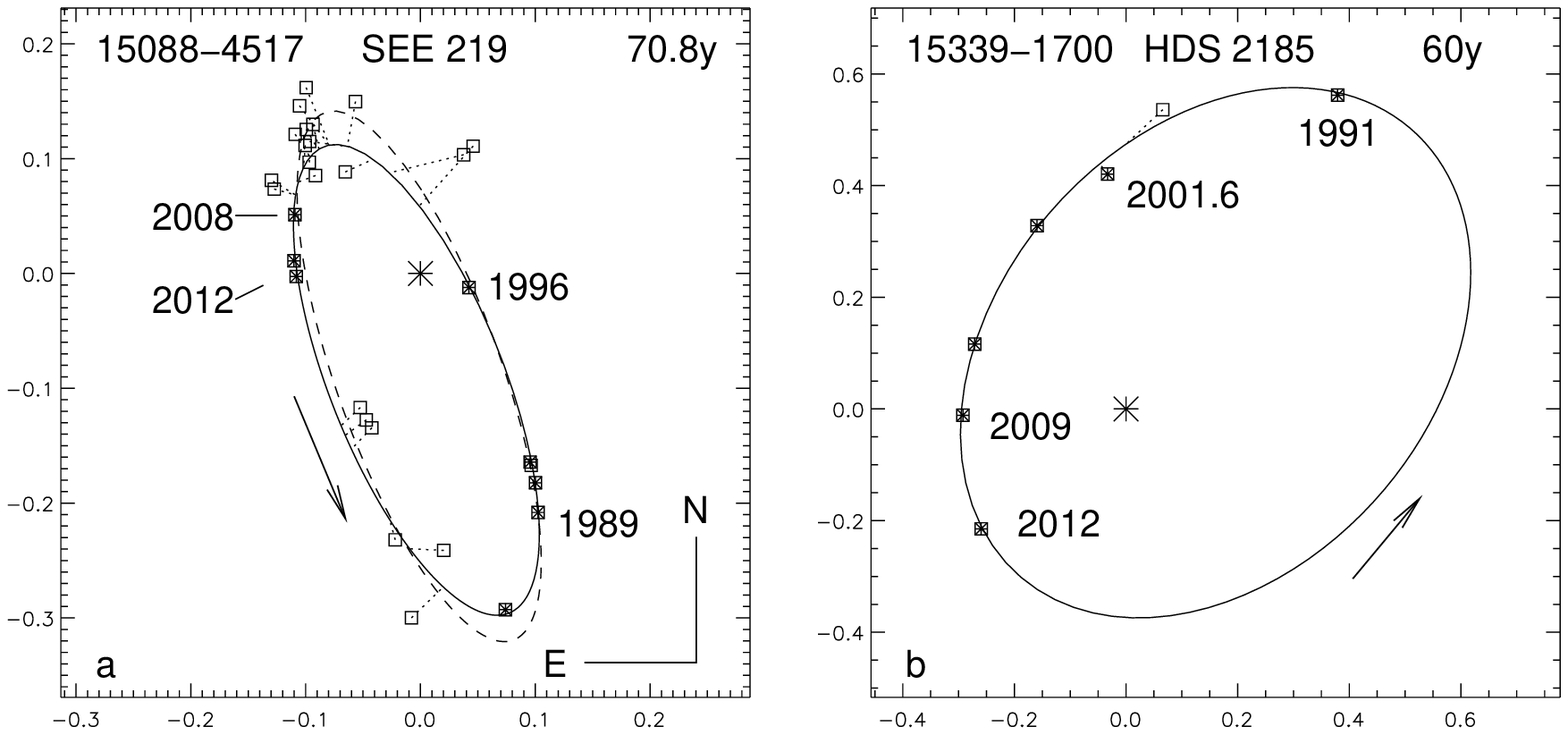}
\caption{New orbits (IV).
\label{fig:4} }
\end{figure*}

{\bf 06359$-$3605 = RST 4816 Ba,Bb}.  The new orbit with 14\,yr period
is radically  different from  the 28.5\,yr orbit  of \citet{Cve2008b},
but closer  to the three solutions proposed  by \citet{Bra2009}. Large
part  of the  orbit remains  unobserved.  This  pair forms  a physical
quadruple with another binary  HIP~31509 = FIN~19~Aa,Ab, also measured
here.

{\bf 07518$-$1354 = BU 101}. This  is a minor revision of the orbit by
\citet{Pbx2000b} needed to reduce  its large residual from our measure
in 2009.  Radial  velocities (RVs) from the above  paper were included
in the  combined orbital  solution, but make  little influence  on the
final   elements   which   are   primarily  constrained   by   speckle
interferometry.  The orbit is now extremely well defined.

{\bf 08270$-$5242 = B 1606}.  The orbit by \citet{Fin1963c} is revised
here using  the three  available speckle measures,  leading to  a more
accurate   period  and   to  the   reduced  orbit   size.   Systematic
over-estimation of the separation by Finsen's visual interferometry is
apparent in Figure~1c.

{\bf  08345$-$3236 =  FIN 335}.  We confirm  and slightly  correct the
orbit by \citet{Sod1999}. The 17.35\,yr period is very well defined now,
but  further  coverage by  speckle  is  still  needed. The  system  is
evolved, judging  from its luminosity, estimated  masses, and spectral
type G5IV.  The speckle $\Delta y$ is very consistent and
preferred to $\Delta$Hp.

{\bf 09173$-$6841= FIN  363 AB} has an unusually  short period of only
3.44\,yr.  The {\it Hipparcos} photometry is doubtful because of close
0\farcs1 separation.

{\bf  11009$-$4030 =  FIN 365}.   The latest  measure  contradicts the
first   orbit  published  in   HTM12,  which   in  fact   predicts  an
unrealistically small  mass sum. We propose here  an alternative orbit
with retrograde motion which  reproduces the non-resolutions by Finsen
in  1963--1966  (Figure~1f)  and  corresponds  to  a  reasonable  mass
sum. Double  lines were noted by  \citet{N04}. The star  appears to be
evolved.   The  {\it Hipparcos}  parallax  could  be  affected by  the
orbital motion unaccounted for in its data reduction.

{\bf 11190$+$1416 = STF 1527} has a long orbital period of 415\,yr, but
it moved  fast through  the periastron in  2009--2012, allowing  us to
compute a  better orbit.  The recent orbit revision  by \citet{Sca2011}
with $P=551$\,yr is not yet included in VB6.

{\bf 11210$-$5429 =  I 879 = $\pi$~Cen}. The  orbit by \citet{Msn1999a}
had to  be revised  using our measures  near the periastron.  The high
eccentricity $e=0.853$ is now well established.

{\bf 12357$-$1650 =  FIN 368 Aa,Ab} is the  first orbit determination.
Speckle measures  of 1989--91 and 2009--11  repeat themselves, hinting
at 20\,yr period.  However, the measure by Mason et al. in 2006.2   does not
fall on  the same  ellipse and had  to be  ignored. It could  refer to
another  star, as  FIN~368 should  have been  unresolved at  that time
according  to our preliminary  orbit, which  also matches  the speckle
non-resolution  in 1976.3  and  the non-resolutions  in 1964--1966  by
Finsen.  An  alternative orbit with $P=10.13$\,yr and  $e=0.9$ can also
be fitted to  the data.  The {\it Hipparcos} measure  on 1991.25 contradicts
speckle interferometry on 1991.39;  it had to be ignored.  \citet{N97}
noted   double   lines   broadened   by   fast   axial   rotation   of
100\,km~s$^{-1}$ and 20\,km~s$^{-1}$.   However, individual 
RVs measured by these authors during 1987--1991 (near the apastron)
do  not show  any systematic  behavior that  could be  related  to the
orbit.  Continued  speckle monitoring will be  critical for confirming
the orbit. The tertiary companion B at 11\farcs8 is physical.


{\bf  13106$-$3128 = RST  1706} is  an example  of a  neglected binary
discovered by R.~A.~Rossiter in 1934  but observed so rarely that only now,
after a nearly full revolution, the first orbit could be proposed.

{\bf 13129$-$5949 = HDS 1850 =  HR 4980} has a tentative edge-on first
orbit with $P=31.6$\,yr.  This  is a chromospherically active G0V dwarf
and a ROSAT X-ray source. There are at least four components in the system:
Aa1,Aa2  is a  double-lined  spectroscopic and  eclipsing binary  with
4.2\,d  period, Aa,Ab  is  the  pair considered  here,  and the  visual
companion B at  25\farcs5 is physical. The orbits  of Aa1,Aa2 and Aa,Ab
may be co-planar. 


{\bf  13169$-$3436   =  I  1567}  has  a   well-established  orbit  by
\citet{Hei1986a}  which is  corrected  here to  better  match the  new
speckle  data.  Heintz  notes that  this pair  is a  ``puzzling case''
because  of some  very discordant  historical measures;  these deviant
points were omitted from Figure~2f and ignored in the calculation.

{\bf 13513$-$2433 = WSI 77  = HD 120690} is a chromospherically active
G5 dwarf within  20\,pc from the Sun.  According  to \citet{Abt06}, it
is also  a single-lined spectroscopic binary with  10.3\,yr period.  We
used RVs from that  work and the average  RV from
\citet{Nidever}  together  with  four  speckle points  for  the  combined
orbital    solution   presented    in    Figure~\ref{fig:WSI77}.   The
spectroscopic elements are $K_1 = 6.06 \pm 0.25$\,km~s$^{-1}$ and $V_0
=   5.38  \pm   0.10$\,km~s$^{-1}$,  the   rms  residual   in   RV  is
0.11\,km~s$^{-1}$.  The node $\omega$  listed in  Table~3 corresponds  to the
primary  component,  therefore $\Omega$  was  chosen  to describe  the
secondary's  relative  motion.   The  pair  was  ``caught''  at  close
separation in 2012.

{\bf  15088$-$4517 =  SEE  219 AB  =  $\lambda$~Lup} is  a B3V  binary
belonging to the Sco-Cen association. A minor revision of the orbit by
\citet{Doc2007d} proposed  here turns it  into a definitive  one, with
both sides of the ellipse now covered by speckle measures and one full
revolution observed (Figure~4a).   The {\it Hipparcos} parallax corresponds to
an uncomfortably large  mass sum.  The true parallax  should be around
6\,mas, matching the distance to the association.

{\bf  15339$-$1700 =  HDS 2185}  has its  first 60\,yr  orbit determined
here, with  nearly half  of it covered  (Figure~4b). The orbit  is still
preliminary.  The speckle measure on 2001.56 was given a lower weight. 




\section{Other results}
\label{sec:other}

\subsection{Astrometric binaries}

The {\it Hipparcos} satellite detected accelerated proper motion $\dot{\mu}$
in some  stars \citep{HIP}. Accelerated  motion is also  inferred from
the difference $\Delta \mu$ between {\it Hipparcos} short-term proper motion
and ground-based  catalogs \citep{MK05,Frankowski}.  These astrometric
observables  do  not  constrain   orbital  periods  and  mass  ratios,
therefore  direct resolution  and follow-up  with adaptive  optics and
speckle  interferometry is  needed.   Such work  has been started  recently
\citep{NICI}.   We continue to  follow  astrometric binaries,
collecting data  for eventual orbit  calculation. The list of  15 such
systems observed  here (including 5  unresolved) is given
in Table~\ref{tab:astrom}.


Astrometric  orbits  for  HIP~38146,   38414,  46396,  and  84924  are
published.   However, they  are inaccurate  and do  not  match speckle
measures  in  position angle.   With  few  more  measures it  will  be
possible  to determine  true visual  orbits, but  so far  this appears
premature.   For  the  remaining  resolved pairs  the  separation  and
parallax are used to  estimate order-of-magnitude orbital periods. The
two    newly   resolved   astrometric    pairs   are    commented   in
Section~\ref{sec:new}.   HIP~64006   shows  elongation  at  $77^\circ$
indicative of  its partial resolution, unless caused  by vibrations or
other artifacts; we do  not consider this resolution secure. Estimated
mass ratio  of HIP~84494 is 0.3, semi-major  axis 0\farcs06, therefore
the companion is below detection limit.

\subsection{Spurious or enigmatic pairs}

\begin{figure}
\epsscale{1.0}
\plotone{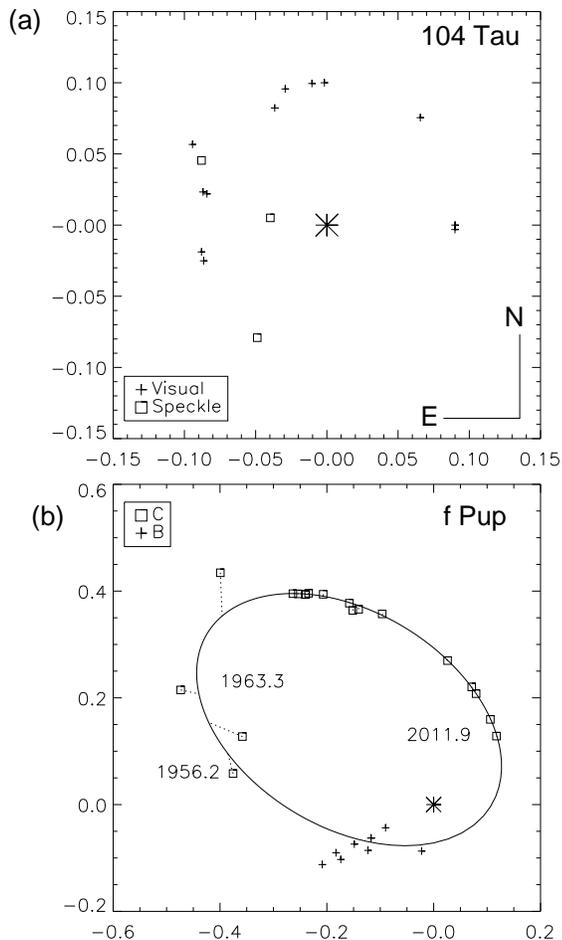}
\caption{Top (a):  motion on the  sky of 104~Tau. Visual  observations are
  plotted as crosses, speckle measures as squares. 
Bottom (b): f~Pup = FIN~324 AC (elliptical orbit connected to squares)
and measures of AB (crosses). The scale on both plots is in arcseconds.
 \label{fig:ghost}}
\end{figure}

Binaries may be unresolved temporarily when their orbital motion makes
them too close.  However, repeated  non-resolutions of a binary with a
short estimated period  put in doubt its veracity.  For example, some
CHARA  speckle  pairs  were  later retracted  by  \citet{Mcalister93}.
Artifacts that may lead to  such spurious discoveries are discussed in
TMH10.

In some  cases, however, binaries were observed  on multiple occasions
by  different people before  disappearing. It  is difficult  to ``write
off'' these binaries  as spurious; rather, they may  point to some new
phenomena. Such ``ghost'' binaries are  brought to light  here. We do
not propose any explanation; the purpose is to attract attention and to
stimulate further collection of data on those stars.
Table~\ref{tab:ghost} lists  close binaries and  sub-systems which were
repeatedly unresolved  in recent speckle  runs at SOAR. It  also gives
$\pi_{\rm  HIP}$,  spectral  type,  and  $V$ magnitude.   Comments  on
individual stars follow.

\begin{figure}
\epsscale{1.0}
\plotone{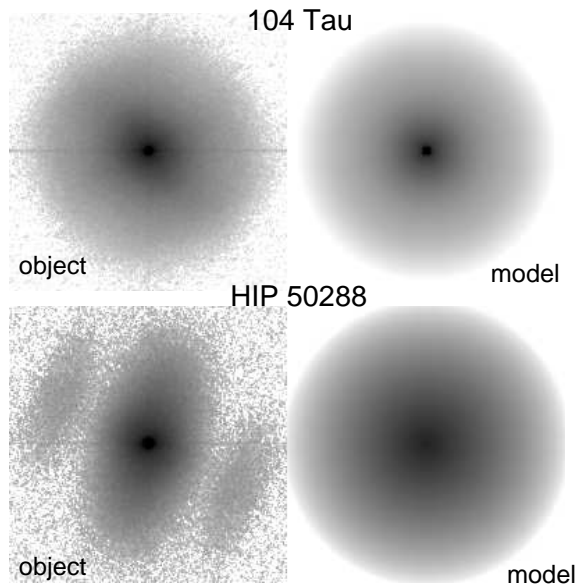}
\caption{The power  spectrum of 104~Tau recorded at  SOAR on 2012.0253
  in the $y$  filter (top) shows  no sign of  fringes (left side  -- data,
  right  side --  their model  for a  single star).   The  gray levels
  display power  on negative logarithmic scale  from $10^{-7}$ (white)
  to $10^{-3}$ (black).
 For  comparison,  the power  spectrum  of  resolved binary  HIP~50288
 (0\farcs035, $\Delta  y = 0.8$)  recorded with the same  equipment on
 2012.0258 is shown in the bottom.
 \label{fig:power}}
\end{figure}

{\bf  05074$+$1839 =  104  Tau} is  a  G4V dwarf  at  16\,pc
resolved into equal components at 0\farcs1 by R.~Aitken in 1912.
The  WDS   contains 16  resolutions  of this  pair
(Figure~\ref{fig:ghost}, top).   Apart from Aitken himself,  it has been
resolved  on  multiple  occasions  by R.H.~Wilson  in  1934--1971,  by
W.~Finsen  (1953--1955) and  by  others, although  in other  instances
those  observers found  it  to  be single.   The  measures plotted  in
Figure~\ref{fig:ghost}a suggest a near-circular orbit seen face-on with a
semi-major axis on the order  of 0\farcs1 or 1.5\,AU.  Assuming a mass
sum of  2\,$M_\odot$, the orbital  period should be around  1.3\,yr; in
fact, two  orbits with periods  1.19\,yr and 2.38\,yr were  published by
\citet{Eggen56}.   This binary should  be an  easy target  for speckle
interferometry  at 4-m  telescopes.   It was  observed  10 times  from
1976.9 to  1980.7 with speckle  and, surprisingly, found  unresolved on
all occasions, excluding any short-period orbits.  Later, however, two
measures were made by the author at 1-m telescope with a phase-grating
interferometer.   The  first resolution  in  1984.8  at 0\farcs04  was
tentative (below the diffraction limit),  but the second one in 1985.7
was secure, being average of two observations.  It was followed by the
speckle resolution at  4-m telescope in 1988.17, after  which the pair
disappeared  again.    It   was  found  unresolved   in  2012
(Figure~\ref{fig:power}).

The star is well studied.   Two statistical surveys of binaries within
25\,pc    consider   it    to   be    single   \citep{DM91,Raghavan10}.
\citet{Heintz84} state that measures cannot be fitted by any orbit and
conclude: ``Although  this alleged visual binary (ADS  3701) has three
published  orbits,  it is  probably  spurious''.  Several  independent
RV  studies  have shown  that  this star  is not  a
spectroscopic binary.  Precise  RVs measured by \citet{Nidever}
are stable  to better than  100\,m~s$^{-1}$ over 388 days,  excluding orbital
periods from one  to two years with high  confidence.  Data with lower
precision  show   a  constant  RV  of  $+21$\,km~s$^{-1}$   over  many  years
\citep{DM91, Abt06, Raghavan10}.

The star  is located about  $1^m$ above the main  sequence, supporting
the thesis of an equal-component binary. If the orbit is seen face-on,
the RV variation  would be small, especially if  the components are of
equal  brightness  (blended lines  move  in  opposite directions,  the
centroid  stays constant).  However,  the lines  in  this star  remain
narrow and the speckle  non-resolution during 3.8\,yr firmly excludes a
face-on  orbit.  Remember  that 104~Tau  is bright  (no identification
errors possible) and that the components are supposedly equally bright,
hence easy to resolve by speckle.

If this  star is  single (as everything  seems to suggest),  we cannot
dismiss   its   multiple   resolutions   with   micrometer,   eyepiece
interferometer, and speckle as  spurious; occasional image doubling (or
at least elongation) must be real.

{\bf 07374$-$3458}  is the  bright  star f  Puppis (HR
2937,  HIP 37096,  HD  61330).   It was  resolved  as 0\farcs2  binary
FIN~324AB   in   1954.31   by  \citet{Fin1956a} using      double-slit
interferometer  at the  0.7\,m Innes  refractor.  The  components were
comparable in  brightness with  $\Delta m$ from  0.3 to  0.6.  Finsen
published 7 mean  positions resulting from 25 nights  (the last one in
1960.26).   His   measures  show  considerable   scatter  (crosses  in
Figure~\ref{fig:ghost}b);  the motion  of  AB looks  erratic rather  than
regular.

Finsen could not resolve  AB since 1960.29, despite repeated attempts.
However, on  1963.305 he found  another companion C at  0\farcs52 with
$\Delta  m =  0.8$.  In  fact W.H.   van den  Bos saw  both companions
earlier and  measured AB  and AC simultaneously  in 1956.2  and 1959.7
\citep{vdB57, vdB61}.  The pair  AC was measured later by R.H.~Wilson,
      {\it  Hipparcos},  and  various  speckle  interferometers.   All
      speckle observations  show no trace of the  subsystem AB, except
      the one on 1989.305 where B looks doubtful and much fainter than
      C \citep{Mason11}.  No trace of B was seen at SOAR in 2008--2011
      (4 speckle runs).

The  binary system  AC  follows  a Keplerian  orbit  with 81\,yr  period
(HTM12) which  was slightly corrected  here using the  latest measure.
This excludes confusion between the companions (i.e. B and C being the same
star).  Besides, both companions were measured concurrently by van den
Bos.

The  companion B  cannot be  real.  The  closest separation  of  AC is
0\farcs09  according to  the orbit  and excludes  any  sub-system with
comparable separation  because such a triple star would  be dynamically
unstable. The fact that the orbit of  AC is known does not allow us to
explain the apparent non-hierarchical configuration by projection.  If
B  were real,  its  orbital period  would  be on  the  order of  10\,yr
(scaling  from the orbit  of AC)  and it  would have  shown up  in our
speckle data. Like 104~Tau, we have here a binary AB which actually is
not a binary -- a ``ghost''.

{\bf  09125$-$4337 =  FIN  317 Aa,Ab}  is  a close  sub-system in  the
2\farcs9  pair AB  = HJ~4188.   After discovery  of Aa,Ab  in  1951 at
0\farcs116, \citet{Fin1951b}  was unable to resolve the  star again on
twelve  occasions till  1968, except  one other  tentative  measure in
1962.  Yet the object was resolved by speckle in 1989.94 at 0\farcs144
and in 2006.18 at 0\farcs123.  Despite orbital period of $\sim 50$\,yr
estimated from  projected separation, the sub-system  was not detected
in  3  runs  at  SOAR  (2009--2012),  while  the  wider  pair  AB  was
measured.  This  may be  yet  another  case  of erratic  measures  and
non-resolutions.

{\bf  15462$-$2804 =  HR~5856 =  HD~140722}.  The  binary  companion B
discovered by S.W.~Burnham  in 1878 moved since then  by $+6^\circ$ in
angle, now at 0\farcs63 separation.  Considering this slow motion, the
pair is  likely much  wider than it  seems, being seen  in projection.
The sub-system  CHR~50 Aa,Ab was discovered  by speckle interferometry
in 1983.42 at 0\farcs20 \citep{McA1987}. A total of four measures are
listed in the  INT4 catalog, the last one  in 2006.19.  Curiously, the
wide pair AB was measured with speckle at 4-m telescopes several times
(in 1985.50,  1989.31, 1991.39) without resolving  the sub-system.  We
measured the AB and found no trace  of CHR~50 in three runs at SOAR in
2009--2012.   The separation of  CHR~50 implies  an orbital  period of
$\sim50$\,yr.  Yet the few speckle resolutions show a fast motion or a
random scatter. If the sub-system CHR~50 were real, we would expect it
to cause some wobble in the motion of AB, but no such signal is seen.

{\bf 15467$-$4314} is a G5V dwarf at 47\,pc. The WDS catalog notes its
resolution in 1926 by Innes at  0\farcs3 and the last measure in 1935.
The separation corresponds to an  orbital period of $\sim 50$\,yr.  Yet
the system was not resolved by  {\it Hipparcos} in 1991.25 and by speckle in
2001.56,  2008.54, and  2012.18.   \citet{N04} found  only a  marginal
variability of RV during 8\,yr.  The binarity is thus questionable.

\section{Conclusions}
\label{sec:concl}

This work provides follow-up measures of close binary stars to be used
in  calculation or  refinement of  their orbits.   Fifteen  orbits are
contributed  to  the  VB6  catalog.  Speckle  interferometry  is  very
efficient. Only a modest investment  of telescope time (few nights per
year  at 4-m  telescopes)  is needed  to  supply good-quality  speckle
measures  for  calculating orbits  of  fast  binaries  and making  the
existing orbits accurate and definitive.  Bright stars can be observed
in twilight or through transparent clouds.

One  class  of objects  to  benefit  from  the speckle  follow-up  are
{\it Hipparcos}  astrometric binaries,  mostly nearby  low-mass  dwarfs. Two
such  stars  are  resolved here  for  the  first  time, few  more  are
measured.  Astrometry  of these and other  binaries requires knowledge
of  their orbits  to disentangle  them from  parallax and  PM.  Future
space astrometric missions like {\it Gaia} will  be too short to do this and
will rely  heavily on  the VB6  catalog.  This is  one more  reason to
follow the  motion of fast  binaries with speckle  interferometry {\em
  now}.

Determination of a large number of orbits is a routine task.  However,
any  large   sample  contains  unusual   or  particularly  interesting
objects. This  might be the  case of ``ghost'' binary  companions that
have  been resolved  several  times, yet  seem  non-existent. Here  we
attract attention  to two such cases,  104~Tau and f~Pup,  and to some
other  visual companions  with seemingly  erratic motion  and frequent
disappearances.   It is  difficult to  accept that  these resolutions,
some  by  very accomplished  observers,  are  all spurious.  Continued
monitoring of such ``ghosts'' is  needed in hope of collecting crucial
observations and eventually explaining this phenomenon.

\acknowledgments   We  thank  the   operators  of   SOAR  D.~Maturana,
S.~Pizarro, P.~Ugarte, A.~Past\'en for their help with labor-intensive
speckle  observations,  B.~Mason  and  W.I.~Hartkopf  --  for  sharing
archival auto-correlation  of FIN~324,  extracting data from  the WDS,
and commenting  the draft  of this paper.   This work used  the SIMBAD
service  operated  by  Centre  des Donn\'ees  Stellaires  (Strasbourg,
France),  bibliographic references from  the Astrophysics  Data System
maintained  by  SAO/NASA,  and  the  Washington  Double  Star  Catalog
maintained at USNO. Comments by anonymous Referee helped to improve the
presentation.
%


{\it Facilities:}  \facility{SOAR}.

\clearpage



\begin{deluxetable}{l l l  ccc  rc cc l r r l }                                                                                                                                
\tabletypesize{\scriptsize}                                                                                                                                                     
\rotate                                                                                                                                                                         
\tablecaption{Measurements of Binary Stars at SOAR         \label{tab:double} }                                                                                                                                                            
\tablewidth{0pt}                                                                                                                                                                
\tablehead{                                                                                                                                                                     
\colhead{WDS} & \colhead{Discoverer} & \colhead{Other} & \colhead{Epoch} & \colhead{Filt} & \colhead{N} & \colhead{$\theta$} & \colhead{$\rho \sigma_{\theta}$} &               
\colhead{$\rho$} & \colhead{$\sigma \rho$} & \colhead{$\Delta m$} & \colhead{[O$-$C]$_{\theta}$} & \colhead{[O$-$C]$_{\rho}$} & \colhead{Reference} \\         
\colhead{(2000)} & \colhead{Designation} & \colhead{name} & +2000 & & & \colhead{(deg)} & (mas) & ($''$) & (mas) & (mag) & \colhead{(deg)} & \colhead{($''$)}                   
& \colhead{VB6 code}  }   
\startdata     
06003$-$3102 &	HU 1399 AB &	HIP 28442 &	11.9353 &	y &	2 &	113.5 &	0.4 &	0.6945 &	0.3 &	1.2   &	$-$1.5 &	0.014 &	Tok2005  \\ 
06359$-$3605 &	FIN 19 Aa,Ab &	HIP 31509 &	11.9353 &	y &	2 &	351.5 &	0.1 &	0.2824 &	0.1 &	1.3   &	0.8 &	$-$0.001 &	Hrt2011d  \\ 
            &	       &	           &	12.1839 &	y &	2 &	350.3 &	0.0 &	0.2887 &	0.3 &	1.2   &	0.2 &	0.000 &	Hrt2011d  \\ 
06359$-$3605 &	RST 4816 Ba,Bb &HIP 31547 &	11.9353 &	y &	2 &	139.0 &	0.1 &	0.1791 &	0.1 &	0.8   &	33.5 &	$-$0.180 &	*Cve2008b  \\ 
            &	       &	           &	12.1839 &	y &	2 &	136.4 &	0.0 &	0.1849 &	0.1 &	0.7   &	31.3 &	$-$0.169 &	*Cve2008b  \\ 
07187$-$2457 &	FIN 313 Aa,Ab &	HIP 35415 &	11.9353 &	y &	2 &	128.8 &	1.4 &	0.1181 &	2.0 &	0.9    &  &   &   \\ 
07187$-$2457 &	TOK 42 Aa,E &	HIP 35415 &	11.9353 &	y &	2 &	88.1 &	1.2 &	0.9454 &	0.4 &	4.4    &  &   &   \\ 
07294$-$1500 &	STF 1104 AB &	HIP 36395 &	11.9353 &	y &	2 &	34.4 &	0.6 &	1.8309 &	0.4 &	1.3 * &	$-$1.8 &	0.200 &	WSI2004a  \\ 
07374$-$3458 &	FIN 324 AC &	HIP 37096 &	11.9353 &	y &	2 &	317.5 &	0.0 &	0.1740 &	0.2 &	1.6   &	1.5 &	$-$0.000 &	Hrt2012a  \\ 
            &	       &	           &	11.9353 &	H$\alpha$ &	2 &	317.6 &	0.2 &	0.1742 &	0.6 &	1.8   &	1.6 &	0.000 &	Hrt2012a  \\ 
07456$-$3410 &	TOK 193 Aa,Ab &	HIP 37853 &	11.9353 &	y &	2 &	72.1 &	1.4 &	0.3098 &	1.6 &	4.0    &  &   &   \\ 
07479$-$1212 &	STF 1146 &	HIP 38048 &	12.1019 &	y &	2 &	342.1 &	0.2 &	1.1457 &	0.6 &	1.4 * &	0.2 &	0.108 &	Nov2007d  \\ 
07490$-$2455 &	TOK 194 &	HIP 38146 &	11.9353 &	H$\alpha$ &	2 &	155.1 &	1.3 &	0.0512 &	3.7 &	1.7   &	 &	&   \\ 
07518$-$1354 &	BU 101 &	HIP 38382 &	12.0254 &	y &	2 &	288.2 &	0.1 &	0.5077 &	0.2 &	0.9   &	$-$1.4 &	$-$0.029 &	*Pbx2000b  \\ 
            &	       &	           &	12.1019 &	y &	2 &	288.4 &	0.1 &	0.5134 &	0.1 &	1.1   &	$-$1.3 &	$-$0.026 &	*Pbx2000b  \\ 
            &	       &	           &	12.1019 &	I &	2 &	288.4 &	0.1 &	0.5135 &	0.2 &	0.9   &	$-$1.4 &	$-$0.026 &	*Pbx2000b  \\ 
07522$-$4035 &	TOK 195 &	HIP 38414 &	11.9354 &	H$\alpha$ &	2 &	168.4 &	2.7 &	0.0442 &	2.5 &	1.0   &	 &	 &  \\ 
07523$-$2626 &	WSI 54 AB &	HIP 38430 &	11.9353 &	y &	3 &	229.2 &	0.7 &	0.0505 &	0.2 &	1.2 :  &  &   &   \\ 
07523$-$2626 &	WSI 54 AC &	HIP 38430 &	11.9353 &	y &	3 &	230.7 &	2.4 &	0.0964 &	6.6 &	1.6 :  &  &   &   \\ 
08221$-$4059 &	HJ 4087 AB &	HIP 41006 &	11.9354 &	y &	2 &	255.3 &	0.2 &	1.4580 &	0.7 &	0.4 * &	0.9 &	0.006 &	USN2002  \\ 
08250$-$4246 &	CHR 226 Aa,Ab &	HIP 41250 &	11.9354 &	y &	2 &	97.9 &	5.1 &	0.0411 &	10.2 &	2.8    &  &   &   \\ 
08250$-$4246 &	RST 4888 Aa,B &	HIP 41250 &	11.9354 &	y &	2 &	104.1 &	0.3 &	0.5372 &	0.1 &	1.1    &  &   &   \\ 
08270$-$5242 &	B 1606 &	HIP 41426 &	12.1019 &	y &	2 &	290.8 &	0.1 &	0.0924 &	0.1 &	1.0   &	$-$37.8 &	$-$0.002 &	*Fin1963c  \\ 
08280$-$3507 &	FIN 314 Aa,Ab &	HIP 41515 &	11.9354 &	y &	2 &	45.6 &	0.1 &	0.1124 &	0.2 &	1.0   &	$-$3.6 &	$-$0.006 &	Hrt2012a  \\ 
08291$-$4756 &	FIN 315 Aa,Ab &	HIP 41616 &	12.1019 &	y &	2 &	203.2 &	0.4 &	0.1026 &	0.1 &	1.6   &	3.7 &	0.020 &	Cve2010e  \\ 
08327$-$5528 &	HDS 1221 Aa,Ab &	HIP 41906 &	12.1019 &	y &	2 &	45.4 &	0.9 &	0.0747 &	1.3 &	2.5    &  &   &   \\ 
08345$-$3236 &	FIN 335 &	HIP 42075 &	11.9354 &	y &	2 &	203.5 &	0.1 &	0.1060 &	0.0 &	0.5   &	$-$13.8 &	0.020 &	*Sod1999  \\ 
08421$-$5245 &	B 1624 &	HIP 42695 &	12.1019 &	y &	2 &	224.2 &	0.2 &	0.2056 &	0.3 &	1.5   &	$-$0.0 &	0.013 &	Hrt2012a  \\ 
            &	       &	           &	12.1019 &	I &	2 &	223.9 &	0.9 &	0.2057 &	0.2 &	1.2   &	$-$0.3 &	0.013 &	Hrt2012a  \\ 
08447$-$5443 &	I 10AB &	HIP 42913 &	12.1019 &	y &	2 &	279.8 &	1.1 &	0.3874 &	0.1 &	3.3   &	0.3 &	$-$0.012 &	Msn2011b  \\ 
09125$-$4337 &	HJ 4188 AB &	HIP 45189 &	12.1839 &	y &	2 &	280.6 &	0.9 &	2.8831 &	0.6 &	0.8 *  &  &   &   \\ 
09173$-$6841 &	FIN 363 AB &	HIP 45571 &	12.1020 &	y &	2 &	189.9 &	0.0 &	0.1107 &	0.0 &	0.8   &	$-$50.9 &	0.016 &	*Sod1999  \\ 
09174$-$7454 &	I 12 AB &	HIP 45581 &	12.1020 &	y &	2 &	260.5 &	0.3 &	0.3122 &	0.2 &	0.9    &  &   &   \\ 
09191$-$4128 &	CHR 239 &	HIP 45705 &	12.0285 &	I &	2 &	344.9 &	0.8 &	0.1013 &	1.0 &	1.7    &  &   &   \\ 
            &	       &	           &	12.0285 &	y &	1 &	344.9 &	0.4 &	0.1050 &	0.4 &	2.4    &  &   &   \\ 
09194$-$7739 &	KOH 83 Aa,Ab &	HIP 45734 &	12.1020 &	y &	3 &	147.2 &	0.6 &	0.1242 &	0.3 &	1.3    &  &   &   \\ 
            &	       &	           &	12.1020 &	I &	2 &	145.8 &	0.8 &	0.1244 &	0.1 &	0.6    &  &   &   \\ 
            &	       &	           &	12.1020 &	R &	2 &	146.5 &	0.5 &	0.1243 &	0.7 &	0.7    &  &   &   \\ 
09243$-$3926 &	FIN 348 &	HIP 46114 &	12.0285 &	y &	2 &	265.2 &	0.2 &	0.1601 &	0.0 &	0.4   &	0.2 &	$-$0.003 &	Msn2010c  \\ 
09275$-$5806 &	CHR 240 &	HIP 46388 &	12.0285 &	y &	2 &	91.2 &	0.3 &	0.0385 &	0.3 &	0.3    &  &   &   \\ 
09276$-$3500 &	B 2215 &	HIP 46396 &	12.0285 &	y &	2 &	37.8 &	0.5 &	0.0337 &	0.4 &	1.5   &	     &	         &	          \\ 
09307$-$4028 &	COP 1 &	HIP 46651 &	12.0285 &	y &	2 &	107.8 &	0.2 &	0.9185 &	0.2 &	1.5   &	$-$2.4 &	$-$0.083 &	Msn2001c  \\ 
09415$-$1829 &	TOK 43 Aa,Ab &	HIP 47537 &	12.0284 &	y &	2 &	26.7 &	0.2 &	0.4309 &	0.5 &	2.0    &  &   &   \\ 
            &	       &	           &	12.0284 &	I &	2 &	26.7 &	0.1 &	0.4305 &	0.2 &	1.7    &  &   &   \\ 
09416$-$3830 &	TOK 198 &	HIP 47543 &	12.0285 &	I &	2 &	46.8 &	5.3 &	0.4767 &	9.5 &	4.7    &  &   &   \\ 
09442$-$2746 &	FIN 326 &	HIP 47758 &	12.0284 &	y &	2 &	21.7 &	0.0 &	0.1393 &	0.2 &	0.3   &	3.1 &	$-$0.013 &	Msn1999c  \\ 
09527$-$7933 &	KOH 86 &	HD 86588 &	12.1020 &	I &	2 &	287.5 &	0.8 &	0.2887 &	1.5 &	1.9    &  &   &   \\ 
10093+2020 &	A 2145 &	HIP 49747 &	12.1840 &	y &	2 &	231.5 &	0.1 &	0.2076 &	0.3 &	0.8   &	$-$10.9 &	0.032 &	Msn2011e  \\ 
10161$-$2837 &	TOK 199 &	HIP 50288 &	12.0258 &	y &	2 &	99.9 &	0.1 &	0.0353 &	0.1 &	0.7    &  &   &   \\ 
10282$-$2548 &	B 199 AC &	HIP 51255 &	12.0258 &	y &	2 &	174.5 &	4.9 &	1.1578 &	4.4 &	3.7    &  &   &   \\ 
10282$-$2548 &	FIN 308 AB &	HIP 51255 &	12.0258 &	y &	2 &	108.9 &	2.7 &	0.1363 &	1.4 &	0.7   &	0.4 &	0.002 &	Hrt2012a  \\ 
10311$-$2411 &	B 201 AB &	HIP 51501 &	12.0258 &	y &	2 &	68.1 &	1.0 &	1.9812 &	1.0 &	4.8 *  &  &   &   \\ 
10361$-$2641 &	BU 411 &	HIP 51885 &	12.0259 &	y &	2 &	306.1 &	0.3 &	1.3329 &	0.4 &	1.1 * &	$-$0.6 &	$-$0.055 &	Sca2001b  \\ 
10370$-$0850 &	A 556 Aa,B &	HIP 51966 &	12.1840 &	y &	2 &	196.8 &	0.3 &	0.8737 &	0.4 &	2.9   &	$-$6.0 &	0.102 &	Pop1978  \\ 
            &	       &	           &	12.1840 &	I &	2 &	196.8 &	0.4 &	0.8738 &	0.2 &	2.2   &	$-$6.0 &	0.102 &	Pop1978  \\ 
10370$-$0850 &	TOK 44 Aa,Ab &	HIP 51966 &	12.1840 &	y &	2 &	58.0 &	1.4 &	0.1510 &	0.6 &	2.8    &  &   &   \\ 
            &	       &	           &	12.1840 &	I &	2 &	58.1 &	0.3 &	0.1509 &	0.1 &	2.3    &  &   &   \\ 
10373$-$4814 &	SEE 119 &	HIP 51986 &	12.1841 &	y &	2 &	272.1 &	0.6 &	0.4361 &	0.8 &	1.7   &	2.4 &	0.006 &	Doc2005f  \\ 
10419$-$7811 &	HDS 1530 &	HIP 52351 &	12.1020 &	I &	2 &	152.2 &	0.5 &	0.1638 &	0.3 &	0.8    &  &   &   \\ 
10465$-$6416 &	FIN 364 AB &	HIP 52701 &	12.1020 &	y &	2 &	151.7 &	0.9 &	0.0737 &	1.6 &	0.7   &	1.5 &	0.012 &	Hrt2012a  \\ 
            &	       &	           &	12.1840 &	y &	3 &	150.6 &	2.2 &	0.0759 &	1.4 &	0.8   &	1.7 &	0.013 &	Hrt2012a  \\ 
            &	       &	           &	12.1840 &	I &	2 &	152.2 &	0.8 &	0.0761 &	0.6 &	0.8   &	3.3 &	0.013 &	Hrt2012a  \\ 
10465$-$6416 &	TOK 45 AC &	HIP 52701 &	12.1020 &	y &	2 &	13.9 &	0.5 &	0.7072 &	1.0 &	4.3    &  &   &   \\ 
            &	       &	           &	12.1840 &	y &	3 &	13.6 &	1.1 &	0.7086 &	4.9 &	4.5    &  &   &   \\ 
            &	       &	           &	12.1840 &	I &	2 &	13.5 &	1.2 &	0.7082 &	0.4 &	3.8    &  &   &   \\ 
11009$-$4030 &	FIN 365 &	HIP 53840 &	12.0258 &	y &	2 &	112.9 &	0.2 &	0.1486 &	0.1 &	0.3   &	$-$7.4 &	0.030 &	*Hrt2012a  \\ 
11014$-$1204 &	HDS 1572 &	HIP 53879 &	12.1023 &	y &	2 &	140.1 &	0.2 &	0.1167 &	0.2 &	1.8    &  &   &   \\ 
11056$-$1105 &	new &	HIP 54214 &	12.1024 &	I &	2 &	60.2 &	2.6 &	0.5923 &	5.6 &	5.4   &	     &	      &	         \\ 
11102$-$1122 &	HDS 1590 &	HIP 54580 &	12.1024 &	y &	2 &	324.9 &	0.1 &	0.1123 &	0.5 &	0.4   &	0.3 &	0.002 &	Hrt2012a  \\ 
11190+1416 &	STF 1527 &	HIP 55254 &	12.1840 &	y &	2 &	208.5 &	0.5 &	0.3186 &	0.1 &	1.2   &	$-$60.1 &	0.158 &	*Pru2009  \\ 
11210$-$5429 &	I 879 &	HIP 55425 &	12.1842 &	y &	2 &	253.9 &	0.3 &	0.1163 &	0.0 &	1.5   &	$-$11.9 &	0.047 &	*Msn1999a  \\ 
11318$-$2047 &	new &	HIP 56245 &	12.1023 &	I &	2 &	65.2 &	2.7 &	1.0501 &	2.3 &	4.9 *  &  &   &   \\ 
11342+1101 &	HDS 1642 &	HIP 56429 &	12.1840 &	y &	2 &	245.6 &	0.4 &	1.1225 &	0.4 &	2.6 *  &  &   &   \\ 
11514+1148 &	HDS 1672 &	HIP 57821 &	12.1840 &	y &	2 &	87.5 &	0.7 &	0.1882 &	0.7 &	3.6    &  &   &   \\ 
            &	       &	           &	12.1840 &	I &	2 &	87.1 &	4.5 &	0.1871 &	3.2 &	3.2    &  &   &   \\ 
11525$-$1408 &	HDS 1676 &	HIP 57894 &	12.1023 &	y &	2 &	126.7 &	0.2 &	0.1902 &	0.2 &	1.4    &  &   &   \\ 
12342$-$1812 &	LV 5 &	HIP 61345 &	12.1024 &	y &	3 &	311.9 &	0.5 &	0.9581 &	0.2 &	2.3    &  &   &   \\ 
12357$-$1650 &	FIN 368 Aa,Ab &	HIP 61463 &	12.1024 &	y &	2 &	107.1 &	0.2 &	0.1307 &	0.2 &	1.7    &0.7  & 0.005  & This work   \\ 
12444+2200 &	HDS 1783 &	HIP 62162 &	12.1843 &	y &	1 &	278.2 &	1.9 &	0.1451 &	1.9 &	3.5    &  &   &   \\ 
            &	       &	           &	12.1843 &	I &	2 &	277.1 &	5.2 &	0.1459 &	1.0 &	2.7    &  &   &   \\ 
12485$-$1543 &	WSI 9109 Aa,Ab &	HIP 62505 &	12.1024 &	y &	3 &	288.5 &	0.7 &	0.0806 &	2.3 &	1.3    &  &   &   \\ 
12533+2115 &	STF 1687 AB &	HIP 62886 &	12.1844 &	y &	2 &	193.6 &	0.7 &	1.1651 &	1.3 &	3.0   &	$-$4.1 &	0.133 &	Hei1997  \\ 
13064+2109 &	COU 11 AB &	HIP 63948 &	12.1843 &	y &	3 &	315.6 &	1.5 &	1.7614 &	1.5 &	4.1 *  &  &   &   \\ 
13081$-$6518 &	CHR 247 Aa,Ab &	HIP 64094 &	12.1844 &	y &	2 &	305.8 &	0.7 &	0.0289 &	0.8 &	0.7    & &  &   \\ 
13106$-$3128 &	RST 1706 &	HIP 64292 &	12.1843 &	y &	2 &	124.4 &	0.5 &	0.1744 &	0.5 &	2.3    & $-$0.5  &  0.001  & This work   \\ 
            &	       &	           &	12.1843 &	I &	2 &	124.4 &	0.2 &	0.1744 &	0.3 &	1.8    &  $-$0.5 &   0.001 & This work   \\ 
13126$-$6034 &	WSI 75 Aa,Ab &	HD 114566 &	12.1844 &	y &	2 &	84.0 &	1.1 &	0.0905 &	1.2 &	3.0    &  &   &   \\ 
13129$-$5949 &	HDS 1850 Aa,Ab &	HIP 64478 &	12.1024 &	y &	2 &	99.3 &	1.9 &	0.3167 &	1.8 &	3.6    & 0.5 & $-$0.002  & This work  \\ 
13137$-$6248 &	HDS 1852 &	HIP 64537 &	12.1844 &	I &	2 &	131.9 &	0.4 &	0.1771 &	0.4 &	0.4    &  &   &   \\ 
13138$-$6335 &	WSI 57 &	HD 114737 &	12.1844 &	y &	2 &	54.0 &	0.7 &	0.1932 &	0.7 &	2.3    &  &   &   \\ 
13147$-$6335 &	MLO 3 Aa,B &	HIP 64624 &	12.1844 &	y &	3 &	38.4 &	2.8 &	1.6923 &	2.6 &	3.5    &  &   &   \\ 
13147$-$6335 &	WSI 58 Aa,Ab &	HIP 64624 &	12.1844 &	y &	3 &	279.9 &	1.9 &	0.2460 &	0.7 &	2.2    &  &   &   \\ 
13169$-$3436 &	I 1567 &	HIP 64804 &	12.1843 &	y &	2 &	135.7 &	0.1 &	0.2811 &	0.1 &	1.2   &	$-$3.5 &	0.058 &	*Hei1986a  \\ 
13175$-$0041 &	FIN 350 &	HIP 64838 &	12.1843 &	y &	2 &	8.0 &	0.1 &	0.1214 &	0.2 &	0.6   &	$-$0.5 &	$-$0.001 &	Hor2011b  \\ 
13275+2116 &	TOK 46 &	HD 117078 &	12.1843 &	I &	3 &	2.9 &	0.7 &	0.1263 &	1.4 &	1.1    &  &   &   \\ 
13317$-$0219 &	HDS 1895 &	HIP 65982 &	12.1843 &	y &	2 &	128.6 &	0.7 &	0.1496 &	0.5 &	1.4   &	0.5 &	0.004 &	Hrt2012a  \\ 
13320$-$6519 &	FIN 369 &	HIP 66005 &	12.1844 &	y &	2 &	317.5 &	0.1 &	0.0737 &	0.2 &	0.7   &	$-$0.5 &	0.001 &	Doc2011c  \\ 
13513$-$2423 &	WSI 77      &	HIP 67620 &	12.1843 &	y &	2 &	52.3 &	0.2 &	0.0261 &	0.5 &	1.2    & $-$4.3 & $-$0.014  & This work  \\ 
            &	       &	           &	12.1843 &	H$\alpha$ &	1 &	54.3 &	0.3 &	0.0392 &	0.3 &	1.7    & $-$2.3 & $-$0.001  & This work  \\ 
13539$-$1910 &	HU 898 &	HIP 67859 &	12.1843 &	y &	2 &	295.9 &	0.2 &	0.3442 &	0.2 &	0.6   &	$-$0.5 &	0.001 &	Doc2012a  \\ 
14150$-$6142 &	COO 167 A,Ba &	HIP 69628 &	12.1844 &	I &	2 &	157.0 &	7.2 &	2.7737 &	3.9 &	2.6    &  &   &   \\ 
14150$-$6142 &	WSI 59 Ba,Bb &	HIP 69628 &	12.1844 &	I &	2 &	244.0 &	5.5 &	0.2133 &	1.4 &	2.2    &  &   &   \\ 
14152$-$6739 &	DON 652 &	HIP 69643 &	12.1844 &	I &	2 &	274.6 &	0.7 &	0.4128 &	1.5 &	2.7    &  &   &   \\ 
            &	       &	           &	12.1844 &	y &	1 &	274.8 &	1.7 &	0.4131 &	1.7 &	3.6    &  &   &   \\ 
15088$-$4517 &	SEE 219 AB &	HIP 74117 &	12.1845 &	y &	2 &	91.4 &	0.0 &	0.1082 &	0.0 &	0.8   &	27.0 &	$-$0.010 &	*Doc2007d  \\ 
15143$-$4242 &	B 1273 Aa,B &	HD 134976 &	12.1845 &	y &	1 &	131.9 &	0.0 &	0.5647 &	0.0 &	1.7 :  &  &   &   \\ 
            &	       &	           &	12.1845 &	I &	2 &	131.9 &	0.2 &	0.5624 &	0.2 &	0.9    &  &   &   \\ 
15143$-$4242 &	WSI 82 Aa,Ab &	HD 134976 &	12.1845 &	y &	1 &	41.7 &	0.0 &	0.0504 &	0.0 &	2.2 :  &  &   &   \\ 
            &	       &	           &	12.1845 &	I &	2 &	45.3 &	2.1 &	0.0507 &	2.1 &	1.4    &  &   &   \\ 
15234$-$5919 &	HJ 4757 &	HIP 75323 &	12.1844 &	y &	2 &	0.8 &	0.1 &	0.8239 &	0.8 &	1.0   &	1.7 &	0.001 &	Hrt2010a  \\ 
15332$-$2429 &	SEE 238 Ba,Bb &	HIP 76143 &	12.1845 &	y &	2 &	122.7 &	0.2 &	0.2558 &	0.2 &	0.5   &	$-$1.5 &	$-$0.024 &	Hei1988d  \\ 
15339$-$1700 &	HDS 2185 &	HIP 76203 &	12.1845 &	y &	2 &	309.6 &	0.3 &	0.3368 &	0.3 &	1.8    & 0.0 & 0.002  & This work  \\ 
15348$-$2808 &	RST 1847 Aa,B &	HIP 76275 &	12.1845 &	I &	2 &	335.0 &	2.8 &	0.9433 &	1.6 &	2.0    &  &   &   \\ 
15348$-$2808 &	TOK 49 Aa,Ab &	HIP 76275 &	12.1845 &	I &	2 &	222.0 &	0.8 &	0.1212 &	0.2 &	2.1    &  &   &   \\ 
15351$-$4110 &	HJ 4786 &	HIP 76297 &	12.1845 &	y &	2 &	275.7 &	0.1 &	0.8242 &	0.1 &	0.6   &	$-$1.0 &	$-$0.002 &	Hei1990c  \\ 
15355$-$4751 &	HDS 2191 &	HIP 76328 &	12.1845 &	I &	2 &	201.4 &	0.4 &	0.2298 &	0.5 &	0.8    &  &   &   \\ 
15428$-$1601 &	BU 35 AB &	HIP 76954 &	12.1845 &	y &	2 &	109.2 &	0.4 &	1.6935 &	0.1 &	1.3 *  &  &   &   \\ 
             &	        &	          &	12.3539 &	y &	2 &	109.4 &	0.2 &	1.6920 &	1.1 &	1.5 *  &  &   &   \\ 
15462$-$2804 &	BU 620 AB &	HIP 77235 &	12.1845 &	I &	2 &	173.4 &	0.1 &	0.6324 &	0.1 &	0.5    &  &   &   \\ 
            &	       &	           &	12.1845 &	y &	1 &	173.5 &	0.0 &	0.6323 &	0.0 &	0.8    &  &   &   \\ 
15471$-$5107 &	B 1790 A,Ba &	HD 140662 &	12.1844 &	I &	2 &	84.0 &	3.4 &	0.4257 &	1.6 &	1.0    &  &   &   \\ 
15471$-$5107 &	WSI 78 Ba,Bb &	HD 140662 &	12.1844 &	I &	2 &	225.5 &	0.2 &	0.0654 &	0.4 &	0.2    &  &   &   \\ 
16048$-$4044 &	I 1284 &	HIP 78765 &	12.3539 &	y &	2 &	234.9 &	0.1 &	0.1224 &	0.1 &	0.7    &  &   &   \\ 
16057$-$3252 &	SEE 264 A,Ba &	HIP 78842 &	12.3539 &	y &	2 &	15.0 &	0.4 &	0.8140 &	0.2 &	1.6   &	1.5 &	0.028 &	Doc2009f  \\ 
            &	       &	           &	12.3539 &	I &	1 &	15.0 &	0.0 &	0.8129 &	0.0 &	1.3   &	1.5 &	0.027 &	Doc2009f  \\ 
16057$-$3252 &	WSI 84 Ba,Bb &	HIP 78842 &	12.3539 &	y &	2 &	163.2 &	0.6 &	0.0979 &	0.4 &	0.0    &  &   &   \\ 
            &	       &	           &	12.3539 &	I &	1 &	162.6 &	0.0 &	0.0973 &	0.0 &	0.0    &  &   &   \\ 
16120$-$1928 &	BU 120 Aa,B &	HIP 79374 &	12.3540 &	y &	3 &	1.4 &	0.3 &	1.3586 &	0.7 &	1.2    &  &   &   \\ 
16120$-$1928 &	CHR 146 Aa,Ab &	HIP 79374 &	12.3540 &	y &	3 &	345.8 &	0.9 &	0.0758 &	3.5 &	2.4    &  &   &   \\ 
16253$-$4909 &	COO 197 Aa,B &	HIP 80448 &	12.1846 &	y &	2 &	95.6 &	5.7 &	2.3138 &	1.4 &	1.5   &	0.8 &	0.029 &	Alz2009b  \\ 
            &	       &	           &	12.1846 &	I &	2 &	95.7 &	3.5 &	2.3083 &	0.1 &	0.9   &	0.9 &	0.023 &	Alz2009b  \\ 
16253$-$4909 &	TOK 50 Aa,Ab &	HIP 80448 &	12.1846 &	y &	2 &	184.8 &	2.9 &	0.2188 &	11.6 &	3.5    &  &   &   \\ 
            &	       &	           &	12.1846 &	I &	2 &	184.6 &	0.5 &	0.2219 &	3.1 &	2.9    &  &   &   \\ 
16385$-$5728 &	RST 869 Aa,B &	HIP 81478 &	12.1846 &	I &	2 &	45.0 &	0.1 &	0.8351 &	0.9 &	4.1    &  &   &   \\ 
            &	       &	           &	12.1846 &	y &	2 &	44.8 &	3.8 &	0.8426 &	6.2 &	3.4    &  &   &   \\ 
16385$-$5728 &	TOK 51 Aa,Ab &	HIP 81478 &	12.1846 &	I &	2 &	59.8 &	2.7 &	0.3327 &	1.3 &	2.7    &  &   &   \\ 
            &	       &	           &	12.1846 &	y &	2 &	59.1 &	4.2 &	0.3311 &	3.8 &	3.8    &  &   &   \\ 
16391$-$3713 &	FIN 340 AB &	HIP 81523 &	12.3539 &	y &	2 &	189.8 &	0.0 &	0.0452 &	0.1 &	0.0   &	15.5 &	$-$0.008 &	Hrt2012a  \\ 
16534$-$2025 &	WSI 86 &	HIP 82621 &	12.3540 &	I &	2 &	109.9 &	3.1 &	0.1851 &	2.9 &	2.9    &  &   &   \\ 
17031$-$5314 &	HDS 2412 Aa,Ab &	HIP 83431 &	12.3540 &	y &	2 &	188.1 &	0.3 &	0.6209 &	0.2 &	3.3    &  &   &   \\ 
17066+0039 &	BU 823 A,Ba &	HIP 83716 &	12.3541 &	y &	2 &	166.9 &	0.6 &	1.0027 &	2.9 &	2.0   &	$-$1.8 &	$-$0.007 &	Hrt2000c  \\ 
            &	       &	           &	12.3541 &	I &	2 &	167.1 &	4.1 &	1.0058 &	9.6 &	1.5   &	$-$1.6 &	$-$0.004 &	Hrt2000c  \\ 
17066+0039 &	TOK 52 Ba,Bb &	HIP 83716 &	12.3541 &	y &	2 &	121.5 &	4.0 &	0.0437 &	0.4 &	0.1    &  &   &   \\ 
            &	       &	           &	12.3541 &	I &	2 &	110.7 &	4.4 &	0.0471 &	10.0 &	0.2    &  &   &   \\ 
17157$-$0949 &	A 2592 A,Ba &	HIP 84430 &	12.3541 &	y &	2 &	122.6 &	0.9 &	0.1754 &	1.6 &	1.4   &	$-$23.6 &	0.002 &	Hei1996c  \\ 
17157$-$0949 &	TOK 53 Ba,Bb &	HIP 84430 &	12.3541 &	y &	2 &	215.9 &	1.8 &	0.0380 &	2.1 &	0.3    &  &   &   \\ 
17195$-$5004 &	FIN 356 &	HIP 84759 &	12.3540 &	y &	1 &	70.7 &	0.0 &	0.0617 &	0.0 &	0.2    &  &   &   \\ 
17248$-$5913 &	I 385 AB &	HIP 85216 &	12.3540 &	y &	2 &	120.1 &	0.6 &	0.3984 &	1.4 &	0.5    &  &   &   \\ 
17248$-$5913 &	I 385 AD &	HIP 85216 &	12.3540 &	y &	2 &	271.7 &	1.6 &	0.2690 &	0.2 &	0.5    &  &   &   \\ 
17297$-$4947 &	I 1323 &	HIP 85610 &	12.3540 &	y &	2 &	105.7 &	0.2 &	0.2598 &	0.1 &	0.5    &  &   &   \\ 
17305$-$1006 &	RST 3978 &	HIP 85675 &	12.3541 &	y &	2 &	72.8 &	0.9 &	0.2048 &	0.6 &	2.3    &  &   &   \\ 
17390+0240 &	WSI 88 &	HIP 86374 &	12.3541 &	y &	2 &	6.7 &	0.6 &	0.1884 &	1.1 &	2.9    &  &   &   \\ 
17415$-$5348 &	HDS 2502 &	HIP 86569 &	12.3540 &	y &	2 &	166.4 &	0.7 &	0.1821 &	0.2 &	0.6    &  &   &   \\ 
17535$-$0355 &	TOK 54 &	V2610 Oph &	12.3541 &	y &	2 &	314.5 &	0.6 &	0.1149 &	0.2 &	0.8    &  &   &   \\ 
17563+0259 &	A 2189 &	HIP 87811 &	12.3541 &	y &	2 &	275.2 &	0.4 &	0.1048 &	0.3 &	1.0   &	$-$6.0 &	0.006 &	Doc2008a  \\ 
17575$-$5740 &	HJ 4992 A,Ba &	HIP 87914 &	12.3540 &	y &	2 &	40.8 &	4.2 &	2.5736 &	3.9 &	2.0    &  &   &   \\ 
            &	       &	           &	12.3540 &	I &	2 &	40.8 &	3.3 &	2.5732 &	3.4 &	1.3    &  &   &   \\ 
17575$-$5740 &	TOK 55 Ba,Bb &	HIP 87914 &	12.3540 &	y &	2 &	160.1 &	1.4 &	0.1485 &	5.8 &	0.4    &  &   &   \\ 
            &	       &	           &	12.3540 &	I &	2 &	161.3 &	4.5 &	0.1392 &	6.2 &	0.4    &  &   &   \\ 
\enddata 
\end{deluxetable}   
                                                                                         

\clearpage



\begin{deluxetable}{l l l   c c c  c c }                                                     
\tabletypesize{\scriptsize}                                                                     
\tablecaption{Unresolved Stars                                                                   
\label{tab:single} }                                                                            
\tablewidth{0pt}                                                                                
\tablehead{                                                                                     
WDS (2000) & \colhead{Discoverer} & \colhead{HIP} & \colhead{Epoch} & \colhead{Filter} &  
\colhead{N} & \multicolumn{2}{c}{5$\sigma$ Detection Limit}   \\             
 & \colhead{Designation} &  & \colhead{+2000} & & &                           
\colhead{$\Delta m (0\farcs15)$}  & \colhead{$\Delta m (1'')$}        \\             
 & \colhead{or other name}  &   & & & & \colhead{(mag)} &  \colhead{(mag)}      
}                                                                                               
\startdata     
05074+1839 &	A 3010 &	HIP 23835 &	12.0253 &	H$\alpha$ &	1 &	3.82 &	4.51  \\ 
            &	       &	           &	12.0253 &	y &	1 &	4.91 &	6.20  \\ 
            &	       &	           &	12.0253 &	V &	1 &	4.62 &	5.56  \\ 
07383$-$2522 &	B 731 &	HIP 37173          &	11.9353 &	y &	2 &	5.01 &	6.85  \\ 
07522$-$4035 &	TOK 195 &	HIP 38414 &	11.9354 &	y &	2 &	4.83 &	6.95  \\ 
08095$-$4720 &	TOK 2 Aa,Ab &	HIP 39953 &	12.1019 &	y &	4 &	5.17 &	6.98  \\ 
09024$-$6624 &	TOK 197 &	HIP 44382 &	12.1020 &	y &	2 &	4.93 &	6.49  \\ 
09380$-$5924 &	HIP 47263 &	HIP 47263 &	12.0285 &	y &	2 &	4.69 &	5.17  \\ 
09416$-$3830 &	TOK 198 &	HIP 47543 &	12.0285 &	y &	1 &	4.57 &	5.08  \\ 
11056$-$1105 &	new     &	HIP 54214 &	12.1024 &	y &	2 &	4.98 &	6.66  \\ 
11126$-$4906 &	HIP 54746 &	HIP 54746 &	12.1842 &	y &	2 &	4.79 &	6.70  \\ 
            &	       &	           &	12.1842 &	I &	2 &	4.04 &	5.72  \\ 
11154$-$5249 &	HIP 54977 &	HIP 54977 &	12.1842 &	y &	2 &	4.02 &	4.75  \\ 
            &	       &	           &	12.1842 &	I &	2 &	3.09 &	4.60  \\ 
11234$-$1847 &	HIP 55598 &	HIP 55598 &	12.1023 &	I &	2 &	3.85 &	5.56  \\ 
11317+1422 &	WSI 9107 Aa,Ab &HIP 56242 &	12.1840 &	y &	2 &	5.00 &	6.15  \\ 
13069$-$3407 &	HIP 64006 &	HIP 64006 &	12.1842 &	I &	2 &	3.97 &	5.73  \\ 
13143$-$5906 &	HIP 64583 &	HIP 64583 &	12.1024 &	y &	2 &	5.22 &	6.67  \\ 
            &	       &	           &	12.1024 &	I &	2 &	3.79 &	6.12  \\ 
13526$-$1843 &	WSI 78 &	HIP 67744 &	12.1843 &	y &	2 &	4.74 &	5.66  \\ 
15168$-$1302 &	CHR 44 &	HIP 74765 &	12.1845 &	y &	2 &	4.64 &	5.54  \\ 
            &	       &	           &	12.1845 &	I &	1 &	3.87 &	5.35  \\ 
15384$-$1955 &	CHR 48 &	HIP 76582 &	12.3539 &	y &	2 &	4.80 &	6.60  \\ 
            &	       &	           &	12.3539 &	I &	2 &	3.97 &	5.86  \\ 
15355$-$1447 &	WRH 20 Aa,Ab &	HIP 76333 &	12.1845 &	y &	2 &	5.05 &	7.03  \\ 
15453$-$5841 &	FIN 234 AB &	HIP 77160 &	12.1844 &	y &	2 &	4.19 &	4.76  \\ 
15467$-$4314 &	I 1276 &	HIP 77282 &	12.1845 &	y &	1 &	4.47 &	5.22  \\ 
            &	       &	           &	12.1845 &	I &	1 &	3.57 &	5.14  \\ 
16057$-$3252 &	SEE 264C &	HIP 78842 &	12.3539 &	y &	1 &	3.31 &	3.63  \\ 
            &	       &	           &	12.3539 &	I &	1 &	3.99 &	5.22  \\ 
16245$-$3734 &	B 868AB &	HIP 80390 &	12.3539 &	y &	2 &	5.02 &	7.10  \\
16534$-$2025 &	WSI 86 &	HIP 82621 &	12.3540 &	y &	2 &	5.11 &	7.25  \\ 
16542$-$4150 &	CHR 252Aa,Ab &	HIP 82691 &	12.3539 &	I &	1 &	3.65 &	5.28  \\ 
16544$-$3806 &	HDS2392 &	HIP 82709 &	12.3539 &	I &	2 &	3.87 &	4.92  \\ 
16571$-$1749 &	HIP 82956 &	HIP 82956 &	12.3540 &	y &	2 &	4.78 &	6.30  \\ 
            &	       &	           &	12.3540 &	I &	2 &	4.06 &	5.89  \\ 
17213$-$5107 &	HIP 84924 &	HIP 84924 &	12.3540 &	y &	2 &	4.90 &	6.29  \\ 
            &	       &	           &	12.3540 &	I &	2 &	3.67 &	5.67  \\ 
\enddata 
\end{deluxetable}


\begin{deluxetable}{l l l rrrrrrrcl}
\tabletypesize{\scriptsize}
\tablewidth{0pt}
\tablecaption{New and Revised Orbital Elements}
\tablehead{
\colhead{WDS} & 
\colhead{Discoverer} &
\colhead{~~~P} & 
\colhead{~a} & 
\colhead{~~i} &
\colhead{~~$\Omega$} &
\colhead{~~T$_o$} &
\colhead{~~e} & 
\colhead{~~$\omega$} &
\colhead{Gr} &
\colhead{Published Orbit} \\
\colhead{(Figure)} &
\colhead{HIP} & 
\colhead{~~~(yr)} & 
\colhead{~($''$)} & 
\colhead{~~($\circ$)} &
\colhead{~~($\circ$)} &
\colhead{~~(yr)} &
\colhead{ } & 
\colhead{~~($\circ$)} &
\colhead{ } & 
\colhead{VB6 Reference} \\
}
\startdata
06359$-$3605 & RST   4816 Ba,Bb &      14.00  &      0.1824 &    111.6  &     289.7  &   1990.036 &      0.577  &     296.1  & 3 & Cve2008b                 \\  
~~~~~~(1a)   &  31547           &  $\pm$0.04  & $\pm$0.0076 & $\pm$1.5  &  $\pm$0.9  & $\pm$0.061 & $\pm$0.031  &  $\pm$1.6  &   &                          \\
             &                  &             &             &           &            &            &             &            &   &                          \\
07518$-$1354 & BU    101        &      23.330 &      0.6179 &     80.82 &     282.65 &   1985.923 &      0.7647 &     253.64 & 1 & Pbx2000b                 \\  
~~~~~~(1b)   &  38382           &  $\pm$0.010 & $\pm$0.0024 & $\pm$0.06 &  $\pm$0.09 & $\pm$0.011 & $\pm$0.0021 & $\pm$ 0.12 &   &                          \\
             &                  &             &             &           &            &            &             &            &   &                          \\
08270$-$5242 & B    1606        &      14.778 &      0.1496 &     59.5  &     96.5   &   1952.54  &      0.337  &     183.8  & 2 & Fin1963c                 \\  
~~~~~~(1c)   &  41426           & $\pm$ 0.077 & $\pm$0.0062 & $\pm$2.5  &  $\pm$2.5  & $\pm$0.46  & $\pm$0.023  & $\pm$ 8.1  &   &                          \\
             &                  &             &             &           &            &            &             &            &   &                          \\
08345$-$3236 & FIN   335        &      17.35  &      0.1445 &     37.5  &      100.1 &   1997.00  &      0.5564 &     217.2  & 3 & Sod1999                  \\  
~~~~~~(1d)   &  42075           &  $\pm$0.05  & $\pm$0.0056 & $\pm$4.5  &  $\pm$3.5  & $\pm$0.16  & $\pm$0.0099 &  $\pm$ 4.2 &   &                          \\
             &                  &             &             &           &            &            &             &            &   &                          \\
09173$-$6841 & FIN  363   AB    &      3.4400 &      0.0894 &    140.2  &     158.3  &   2009.892 &      0.4505 &     118.2  & 2 & Sod1999                  \\  
~~~~~~(1e)   &   45571          &  $\pm$0.0049& $\pm$0.0014 & $\pm$2.1  &  $\pm$3.1  & $\pm$0.020 & $\pm$0.0125 &  $\pm$3.5  &   &                          \\
             &                  &             &             &           &            &            &             &            &   &                          \\
11009$-$4030 & FIN   365        &      27.09  &      0.1523 &    105.1  &      102.6 &   1993.15  &      0.188  &      96.4  & 4 & Hrt2012a                 \\  
~~~~~~(1f)   &   53840          &  $\pm$0.37  & $\pm$0.0047 & $\pm$1.0  &  $\pm$1.5  & $\pm$0.31  & $\pm$0.044  &  $\pm$ 6.0 &   &                          \\
             &                  &             &             &           &            &            &             &            &   &                          \\
11190$+$1416 & STF  1527        &     415.0   &      2.225  &     55.6  &     188.0  &   2010.503 &      0.8511 &       0.50 & 3 & Pru2009                  \\  
~~~~~~(2a)   & 55254            &  $\pm$15.9  & $\pm$0.045  & $\pm$0.5  &  $\pm$0.5  & $\pm$0.042 & $\pm$0.0033 & $\pm$ 0.59 &   &                          \\
             &                  &             &             &           &            &            &             &            &   &                          \\
11210$-$5429 & I     879        &      39.00  &      0.2263 &     19.4  &     327.8  &   2010.410 &      0.8530 &     340.3  & 2 & Msn1999a                 \\  
~~~~~~(2b)   &  55425           &  $\pm$0.19  & $\pm$0.0011 & $\pm$ 4.9 &  $\pm$ 3.9 & $\pm$0.037 & $\pm$0.0040 &  $\pm$ 4.0 &   &                          \\
             &                  &             &             &           &            &            &             &            &   &                          \\
12357$-$1650 & FIN   368  Aa,Ab &      20.07  &      0.1416 &    100.5  &     116.4  &   1985.52  &      0.33   &     260.0  & 4 & First orbit              \\
~~~~~~(2c)   &   61463          &  $\pm$0.10  & $\pm$0.0009 & $\pm$0.8  &  $\pm$0.6  & $\pm$0.09  &      *      &       *    &   &                          \\
             &                  &             &             &           &            &            &             &            &   &                          \\
13106$-$3128 & RST  1706        &      91.0   &      0.463  &    111.9  &      63.4  &   1999.66  &      0.532  &     168.6  & 4 & First orbit              \\
~~~~~~(2d)   &   64292          &        *    & $\pm$0.014  & $\pm$1.0  &  $\pm$1.7  &  $\pm$0.39 & $\pm$0.0035 & $\pm$ 2.6  &   &                          \\
             &                  &             &             &           &            &            &             &            &   &                          \\
13129$-$5949 & HDS  1850 Aa,Ab  &      31.60  &      0.3136 &     90.0  &     278.8  &   2005.26  &      0.340  &      50.0  & 5 & First orbit              \\
~~~~~~(2e)   &   64478          &  $\pm$0.80  & $\pm$0.0070 &      *    &  $\pm$0.2  &  $\pm$0.11 & $\pm$0.016  &       *    &   &                          \\
             &                  &             &             &           &            &            &             &            &   &                          \\
13169$-$3436 & I     1567       &      41.07  &      0.3292 &    120.2  &     145.8  &   2006.456 &      0.4582 &     275.4  & 2 & Hei1986a                 \\  
~~~~~~(2f)   &    64804         &  $\pm$0.19  & $\pm$0.0017 & $\pm$0.66 &  $\pm$ 0.5 & $\pm$0.042 & $\pm$0.0042 &  $\pm$ 0.8 &   &                          \\
             &                  &             &             &           &            &            &             &            &   &                          \\
13513$-$2433 & WSI  77          &      10.485 &      0.2827 &     96.4  &     351.3  &   2009.218 &      0.3462 &     137.5  & 2 & First, combined    \\  
~~~~~~(3)    &   67620          &  $\pm$ 0.06 & $\pm$0.0014 & $\pm$0.2  &  $\pm$0.3  & $\pm$0.028 & $\pm$0.0080 & $\pm$ 1.4  &   &                          \\
             &                  &             &             &           &            &            &             &            &   &                          \\
15088$-$4517 & SEE   219  AB    &      70.8   &      0.2597 &     71.59 &      26.64 &   1997.907 &      0.6283 &     299.9  & 2 & Doc2007d                 \\
~~~~~~(4a)   &    74117         &  $\pm$0.8   & $\pm$0.0021 & $\pm$0.41 &  $\pm$0.61 & $\pm$0.105 & $\pm$0.0090 &  $\pm$1.1  &   &                          \\
             &                  &             &             &           &            &            &             &            &   &                          \\
15339$-$1700 & HDS  2185        &      60.0   &      0.5391 &     40.0  &     324.0  &   2011.53  &      0.3681 &     152.3  & 5 & First orbit              \\
~~~~~~(4b)   &   138648         &       *     & $\pm$0.0018 &      *    &  $\pm$1.0  & $\pm$0.02  & $\pm$0.0030 &  $\pm$1.0  &   &                          \\
             &                  &             &             &           &            &            &             &            &   &                          \\
\enddata
\end{deluxetable}

\clearpage

\begin{deluxetable}{l r c l ccc  c cc c}
\tabletypesize{\scriptsize}
\tablewidth{0pt}
\tablecaption{Parameters of Orbital Pairs}
\tablehead{
\colhead{WDS} & 
\colhead{HIP} & 
\colhead{$\pi_{\rm HIP}$} & 
\colhead{Sp.} &
\colhead{$V$} &
\colhead{$\Delta$Hp } &
\colhead{$\Delta y$ } &
\colhead{$P$} & 
\colhead{$M_1$} &
\colhead{$M_2$} &
\colhead{$\pi_{\rm dyn}$ } \\
 & & 
\colhead{~~~(mas)} &
\colhead{type} & (mag) & (mag) & (mag) & 
\colhead{~~~(yr)} &  
\colhead{($M_\odot$)} &
\colhead{($M_\odot$)} &
\colhead{~~~(mas)} 
}
\startdata
06359$-$3605 & 31547   & 25.39$\pm$0.43  &    G1V    &    7.23 & 0.83  &   0.8   & 14.0   &   1.08 & 0.80  & 25.5 \\  
07518$-$1354 & 38382   & 60.59$\pm$0.59  &    G0V    &    5.16 & 0.90  &   1.0   & 23.3   &   1.16 & 0.86  & 59.8 \\          
08270$-$5242 & 41426   & 18.50$\pm$0.44  &    F5V    &    6.50 & 0.87  &   1.0   & 14.8   &   1.61 & 1.28  & 17.5 \\  
08345$-$3236 & 42075   & 14.86$\pm$0.57  &    G5IV   &    6.43 & 0.38: &   0.5   & 17.35  &   1.71 & 1.54  & 14.6 \\  
09173$-$6841 & 45571   & 30.64$\pm$0.70  &    F5V    &    5.40 & 1.37: &   0.8   & 3.44   &   1.63 & 1.37  & 27.2 \\  
11009$-$4030 & 53840   & 15.42$\pm$0.42  &    F7V    &    6.79 &\ldots &   0.3   & 27.1   &   1.75 & 1.65  & 11.2 \\  
11190$+$1416 & 55254   & 32.52$\pm$1.39  &    F9V    &    6.95 & 1.08  &   1.1   & 415    &   0.98 & 0.61  & 34.2 \\  
11210$-$5429 & 55425   &  9.12$\pm$0.34  &    B9V    &    3.90 & 1.49  &   1.5   & 39.0   &   6.43 & 3.68  & 9.1  \\  
12357$-$1650 & 61463   & 13.56$\pm$0.76  &    F3IV   &    6.70 & 3.10: &   1.7   & 20.1   &   1.79 & 1.23  &13.3  \\  
13106$-$3128 & 64292   & 19.27$\pm$1.23  &    K0V    &    9.10 & 1.93  &   1.9   & 91.0   &   0.90 & 0.71  &19.6  \\  
13129$-$5949 & 64478   & 23.72$\pm$0.60  &    G0V    &    6.20 & 3.32  &   3.6   & 31.6   &   1.56 & 0.79  &23.6  \\  
13169$-$3436 & 64804   & 23.18$\pm$1.10  &    G5V    &    8.06 & 1.00  &   1.1   & 41.1   &   0.96 & 0.84  &22.7  \\  
13513$-$2433 & 67620   & 51.35$\pm$0.45  &    G5V    &    6.45 &\ldots &   3.5   & 10.5   &   0.99 & 0.63  &50.3  \\  
15088$-$4517 & 74117   & 4.20 $\pm$0.66  &    B3V    &    4.04 & 0.80  &   0.8   & 70.8   &   8.14 & 5.84  & 6.3  \\  
15339$-$1700 & 76203   & 25.85$\pm$0.94  &    G9IV   &    8.14 & 1.74  &   1.8   & 60.0   &   0.90 & 0.72  &29.9  \\  
\enddata
\end{deluxetable}


\begin{deluxetable}{l r l c c c l }
\tabletypesize{\scriptsize}
\tablewidth{0pt}
\tablecaption{Astrometric Binaries \label{tab:astrom}}
\tablehead{
\colhead{WDS} & 
\colhead{HIP} & 
\colhead{Discoverer} & 
\colhead{$\rho$} & 
\colhead{$\Delta V$} & 
\colhead{$P$} &
\colhead{Comment} \\
& & \colhead{code} & 
\colhead{($''$)} &
\colhead{(mag)} &
\colhead{(yr)} & 
}
\startdata
07456$-$3410 & 37853 &TOK 193 & 0.310 & 4.0 & 10? & $\dot{\mu}$, $\Delta \mu$, SB \\
07490$-$2455 & 38146 &TOK 194 & 0.051 & 1.7 & 2.4 & Orbit \citep{Gln2007} \\
07522$-$4035 & 38414 &TOK 195 & 0.044 & 1.0 & 7.0 & Orbit \citep{Jnc2005}, SB1  \\
09191$-$4128 & 45705 &CHR 239 & 0.105 & 2.4 & 10? & $\dot{\mu}$, $\Delta \mu$  \\
09275$-$5806 & 46388 &CHR 240 & 0.039 & 0.3 & 4?  & $\dot{\mu}$ in HIP2 \\
09276$-$3500 & 46396 &B 2215  & 0.034 & 1.5 & 1.97& Orbit \citep{HIP} \\
09416$-$3830 & 47543 & \ldots & \ldots& \ldots & ?&  $\dot{\mu}$ in HIP2 \\
11056$-$1105 & 54214 &new     & 0.592 & 5.4 & 30  & Orbit \citep{Gon2000} \\
11234$-$1847 & 55598 & \ldots & \ldots& \ldots & ?&  $\dot{\mu}$, $\Delta \mu$, SB \\
11318$-$2047 & 56245 &new     & 1.058 & 4.9 & 200? &  $\Delta \mu$ \\
13069$-$3407 & 64006 & new?   & 0.025?& ?   & 1?  &  Elongation? $\dot{\mu}$, $\Delta \mu$ \\
13518$-$2423 & 67620 & WSI 77 & 0.283 & 0.5 & 10.5 & $\Delta \mu$, orbit here, SB1 \\
16534$-$2025 & 82621 & WSI 86 & 0.185 & 2.9 & 10?  &  $\dot{\mu}$, $\Delta \mu$ \\
16571$-$1749 & 82956 & \ldots & \ldots& \ldots & ? & $\dot{\mu}$, $\Delta \mu$, SB \\
17213$-$5107 & 84924 & \ldots & \ldots&  \ldots &  3.94 &   Orbit \citep{HIP}, SB \\
\enddata
\end{deluxetable}


\begin{deluxetable}{l r ccc l l}
\tabletypesize{\scriptsize}
\tablewidth{0pt}
\tablecaption{Spurious or Enigmatic Binaries \label{tab:ghost}}
\tablehead{
\colhead{WDS} & 
\colhead{HIP} & 
\colhead{$\pi_{\rm HIP}$} & 
\colhead{Sp.} &
\colhead{$V$} &
\colhead{Discoverer} &
\colhead{Comment}  \\
 & & \colhead{~~~(mas)} &
\colhead{type} & (mag) & \colhead{code} & 
}
\startdata
 05074$+$1839 &	23835   & 64.8$\pm$0.3 & G4V  &	5.01 & A 3010        & 104 Tau, ADS 3701 \\
 07374$-$3458 & 37096   & 9.1$\pm$0.4  & B8IV & 4.52 & FIN 324 AB    & AC orbit   in HTM12 \\
 09125$-$4337 & 45189   & 4.7$\pm$0.5  & B8V  & 5.56 & FIN 317 Aa,Ab  & AB is HJ 4188 at 2\farcs9  \\
 15462$-$2804 & 77235   & 14.1$\pm$1.1  & F2IV & 6.51 & CHR 50 Aa,Ab    & AB is BU\,620 at 0\farcs63  \\
 15467$-$4314 & 77282   & 21.4$\pm$1.0  & G5V & 8.08  & I 1276         & Spurious? \\
\enddata
\end{deluxetable}

\end{document}